\documentclass[11pt,preprint]{aastex}
\newcommand{\HII}{\ion{H}{2}~}
\newcommand{\HI}{\ion{H}{1}~}
\newcommand{\etal}{et al.\ }
\newcommand{\etalb}{et al.}
\newcommand{\NGST}{{\it NGST}}
\newcommand{\vzrms}{v_{\rm rms,0}}
\newcommand{\vrms}{v_{\rm rms}}
\newcommand{\beq}{\begin{equation}}
\newcommand{\eeq}{\end{equation}}

\begin{document}
\title{Constraints on Warm Dark Matter from Cosmological Reionization}

\author{Rennan Barkana}
\affil{Canadian Institute for Theoretical Astrophysics, 60 St. George 
Street, Toronto, Ontario, M5S 3H8, CANADA}
\email{barkana@cita.utoronto.ca}

\and

\author{Zolt\'an Haiman\altaffilmark{1} and Jeremiah P. Ostriker}
\affil{Princeton University Observatory, Ivy Lane, Princeton, NJ
08544}

\altaffiltext{1}{Hubble Fellow}

\begin{abstract}

We study the constraints that high-redshift structure formation in the
universe places on warm dark matter (WDM) dominated cosmological
models. We modify the extended Press-Schechter formalism to derive the
halo mass function in WDM models. We show that our predictions agree
with recent numerical simulations at low redshift over the halo masses
of interest. Applying our model to galaxy formation at high redshift,
we find that the loss of power on small scales, together with the
delayed collapse of low-mass objects, results in strong limits on the
root-mean-square velocity dispersion $\vzrms$ of the WDM particles at
redshift zero. For fermions decoupling while relativistic, these
limits are equivalent to constraints on the mass $m_X$ of the
particles. The presence of a $\approx 4\times 10^9~{\rm M_\odot}$
supermassive black hole at redshift 5.8, believed to power the quasar
SDSS 1044-1215, implies $m_X\ga 0.5$ keV (or $\vzrms \la 0.10$ km/s),
assuming that the quasar is unlensed and radiating at or below the
Eddington limit. Reionization by redshift 5.8 also implies a limit on
$m_X$. If high-redshift galaxies produce ionizing photons with an
efficiency similar to their redshift-three counterparts, we find $m_X
\ga 1.2~$keV (or $\vzrms \la 0.03$ km/s). However, given the
uncertainties in current measurements from the proximity effect of the
ionizing background at redshift three, values of $m_X$ as low as
$0.75~$keV (or $\vzrms = 0.06$ km/s) are not ruled out. The limit
weakens further to $m_X\ga0.4~$keV (or $\vzrms \la 0.14$ km/s), if,
instead, the ionizing-photon production efficiency is ten times
greater at high redshift, but this limit will tighten considerably if
reionization is shown in the future to have occurred at higher
redshifts. WDM models with $m_X\la1~$keV (or $\vzrms \ga 0.04$ km/s)
produce a low-luminosity cutoff in the high-redshift galaxy luminosity
function which is directly detectable with the {\it Next Generation
Space Telescope},\, and which serves as a direct constraint on $m_X$.

\end{abstract}


\section{Introduction}

\label{sec:Intro}

The currently favored model of hierarchical galaxy formation in a
universe dominated by cold dark matter (CDM) has been very successful
in matching observations of the density distribution on large scales.
These successes include the properties of galaxy clusters, galaxy
clustering on large scales, the statistics of the Lyman-$\alpha$
forest, and the temperature anisotropy of the cosmic microwave
background. However, recently some small-scale shortcomings of this
model have appeared \citep[see][for a recent review]{sk00}. CDM models
predict dense, cuspy dark matter halo profiles \citep{NFW,
moore98,swamy,klypin00} which are not apparent in the mass
distribution derived from measurements of the rotation curves of dwarf
galaxies \citep[e.g.,][]{deBlok97,salucci}, although observational
and modeling uncertainties may preclude a firm conclusion at present
\citep{Bosch00,swaters00}. Dense dark matter halos may contradict
observational indications of fast-rotating bars in galaxies \citep[but
see \citealt{TOst99} for a counterargument]{sellwood1,sellwood2}.
Halos with lower central densities than in CDM also appear necessary
to explain the observed normalization of the Tully-Fisher relation
\citep[but see the reassessment by \citealt{eke01}]{navs00,momao}. In
addition, the abundance of satellites and dwarf galaxies in the Local
Group is apparently lower by an order of magnitude than the abundance
of corresponding halos in numerical simulations of CDM \citep{klypin,
moore}. Finally, the halo profiles expected in CDM may conflict with
observed lensing statistics \citep{Li,keeton}.

Although the significance of these discrepancies is still disputed,
and astrophysical solutions involving feedback may still be possible,
the accumulating tension with observations has focused attention on
solutions involving the particle properties of dark matter.
Self-interacting dark matter particles can heat the low-entropy
particles which would otherwise create a dense core, and result in
flatter halo profiles \citep[but see \citealt{escudeSIDM, koch00,
yoshida00} and \citealt{oleg00} for possible conflicts of the simplest
models with observations]{ss00, burkert00, firmani00,
dave00}. Self-interacting dark matter particles in the initial state
of a cold Bose-Einstein condensate, with a repulsive interaction added
to gravity, can also suppress halo cores \citep{goodman,peebles}.
Alternatively, even with no interaction other than gravity, the
quantum-mechanical wave properties of ultra-light ($\sim 10^{-22}$ eV)
dark matter particles stabilize gravitational collapse, providing halo
cores and sharply suppressing the small-scale linear power spectrum
\citep[``Fuzzy CDM'';][]{fuzzy}. Finally, a resurrection of warm dark
matter (WDM) models \citep{joel1, joel2, pierp} has recently been
proposed \citep*{colin00, larsen, Bode}. By design, a common feature
of models that attempt to solve the apparent ``small-scale'' problems
of CDM is the reduction of power on small scales.  In the CDM
paradigm, structure formation proceeds ``bottom-up'', i.e., the
smallest objects collapse first, and they subsequently merge together
to form larger objects. It then follows that the loss of small-scale
power modifies structure formation most severely at the highest
redshifts; in particular, the number of self-gravitating objects at
high redshift is reduced.

A strong reduction in the abundance of high-redshift objects would,
however, be in conflict with the observed reionization of the
universe.  The lack of a Gunn-Peterson trough in the spectra of
high-redshift quasars implies that the hydrogen in the intergalactic
medium (IGM) was highly ionized throughout the universe. In
particular, the spectrum of the bright quasar SDSS 1044-1215 at
redshift $z=5.8$ \citep{f00} shows transmitted flux short-ward of the
Ly$\alpha$ wavelength, implying that reionization was complete by
$z=5.8$ Note that recent observations of small-scale anisotropies of
the cosmic microwave background radiation \citep{boomerang, maxima}
can be used to also place an upper limit of $z\la 30$ on the
reionization redshift \citep{Zaldar}. The most natural explanation for
reionization is photo-ionizing radiation produced by an early
generation of stars and quasars; recent calculations of structure
formation in $\Lambda$CDM find that reionization should naturally
occur at $z\sim 7$--12 \citep[and for a recent review of reionization,
see \citealt{me01}]{hl97, hl98, go97, chiu, g00}. In these models, the
sources of reionization reside in halos that have masses in the range
corresponding to dwarf galaxies --- the mass scale on which power
needs to be reduced relative to CDM models.

In this paper, we examine the constraints that arise from the
reionization of the universe.  We focus on WDM models, although
similar constraints would apply to other modifications of the CDM
paradigm that reduce the small-scale power. Our goal is to quantify
the allowed mass range of WDM particles, given the evidence that the
universe was reionized by a redshift $z>5.8$.  The rest of this paper
is organized as follows. In \S~2 we review the basic properties of WDM
models.  In \S~3 we describe our methods of modeling the formation and
abundance of WDM halos, as well as reionization.  In \S~4 we present
our main results, namely, the limits on the WDM particle
mass. Finally, in \S~5, we summarize our conclusions and the
implications of this work.  Unless stated otherwise, in the case of
CDM, we adopt the $\Lambda$CDM parameters $\Omega_0=0.3$,
$\Omega_\Lambda=0.7$, and $\Omega_{\rm b}=0.045$ for the energy
density ratios relative to the critical density of matter, vacuum, and
baryons, respectively. We also assume a Hubble constant $H_0=100\,
h~{\rm km~s^{-1}~Mpc^{-1}}$ with $h=0.7$, and a primordial scale
invariant ($n=1$) power spectrum with $\sigma_8=0.9$, where $\sigma_8$
is the root-mean-square amplitude of mass fluctuations in spheres of
radius $8\ h^{-1}$ Mpc. These parameter values are based primarily on
observations of cosmological expansion and large-scale structure
\citep[summarized, e.g., in][]{concord} and degree-scale temperature
anisotropy measurements in the cosmic microwave background
\citep{boomerang,maxima}. If CDM is replaced by WDM with
similar cosmological parameters, the resulting model (in which the
contribution of WDM to $\Omega$ equals $\Omega_0-\Omega_{\rm b}$) is
termed $\Lambda$WDM.

\section{Warm Dark Matter Models}

\label{sec:WDM}

In this paper we focus on WDM models, which attempt to improve over
CDM by reducing small-scale power.  The WDM is assumed to be composed
of particles of about a keV mass (rather than a GeV as often assumed
for CDM). In this case, the particles are not completely cold: thermal
velocities on the order of 1 km/s at the time of reionization produce
free streaming whereby particles stream out of overdense regions. This
smoothes out small regions, leading to a small-scale cutoff in the
linear power spectrum, and it also acts similarly to pressure and
substantially delays the formation of small halos.

We assume that the WDM is composed of particles of mass $m_X$ which
decouple in the early universe while relativistic and in thermal
equilibrium \citep[see, e.g.,][]{text}. The initial Fermi-Dirac
distribution function for the momentum $p$ is given by
$\left\{\exp\left[(p c)/(k_B T_X) \right]+ 1\right\}^{-1}\,$, in
terms of the temperature $T_X$. Since both relativistic and
non-relativistic momenta redshift as $1/a$, where $a$ is the scale
factor, the distribution function remains self-similar as the universe
expands, with the effective temperature $T_X$ (which equals the actual
temperature as long as the WDM is relativistic) also redshifting as
$1/a$. At WDM decoupling, $T_X$ equals the photon temperature $T$, but
later annihilating species transfer their entropy to the photons (and
to other particles in thermal equilibrium with the photons) but not to
the WDM. As a result, today $T_X/T = \left[ (43/11)/ g_{\rm dec}
\right]^{1/3}$, where $g_{\rm dec}$ is the effective number of
relativistic species present at WDM decoupling, and 43/11 is the
number today, assuming three massless neutrino species. The present
number density $n_X$ of WDM relative to that of photons equals
$(T_X/T)^3g_X/2$, where $g_X$ is the effective number of degrees of
freedom of WDM. Bosonic degrees of freedom contribute unity, and
fermionic ones contribute 7/8 to $g_{\rm dec}$ (which measures entropy
density) and 3/4 to $g_X$ (which measures number density). Thus, the
usual assumption of a fermionic spin-$1/2$ particle implies $g_X=1.5$.

To produce a given contribution $\Omega_X$ to the cosmological
critical density, the required mass is determined by $m_X n_X \propto
\Omega_X h^2$. The above distribution function at low redshift has the
form $\left\{ \exp\left[ v/v_0 \right]+ 1\right \}^{-1}\, ,$ where we
used the fact that $p=m_X v$ (with $v$ the physical, peculiar
velocity) after the WDM particles become non-relativistic. The
root-mean-square velocity is $\vrms=3.597\, v_0$, where $m_X v_0 = k_B
T_X/c \propto (n_X/ g_X)^{1/3}$. These relations can be combined to
yield a formula relating $\vrms$ to $m_X$, \beq \vrms(z)= 0.0437 \left(
1+z \right) \left( \frac{ \Omega_X h^2} {0.15}
\right)^{1/3} \left( \frac{g_X} {1.5} \right)^{-1/3} \left( \frac{m_X}
{\rm 1~keV} \right)^ {-4/3}~{\rm km~s^{-1}}\ . \eeq Although not
directly relevant to the astrophysics of WDM, a fundamental aspect of
the particle physics is the required value of \beq g_{\rm dec} = 766
\left( \frac{ \Omega_X h^2} {0.15} \right)^{-1} \left( \frac{g_X}
{1.5} \right) \left( \frac{m_X} {\rm 1~keV} \right)\ . \eeq Since this
is larger by a factor of $\sim 7$ than the number of degrees of
freedom in the standard model, WDM requires physics beyond the
standard model \citep[see, e.g., the discussion in][]{Bode}. Other
alternatives for the origin of the velocities are possible (e.g.,
non-thermal production of weakly interacting massive particles, see
\citealt{Lin01}); the critical element is the value of $\vrms$, which
can always be parameterized as $\vrms= (1+z) \, \vzrms$. As noted
above, for the fiducial model considered here, $\vzrms \approx
0.05~{\rm km~s^{-1}}$.

The growth of linear perturbations in WDM models can be calculated by
solving the Einstein-Boltzmann equations. The resulting linear power
spectrum can be written as the CDM power spectrum \citep[which we
compute using][]{eh99} times the square of a transfer function. For
$z=0$, the transfer function is \citep{Bode} \beq T_X(k)=\left( 1+
(\epsilon k R_c^0)^{2 \nu} \right) ^{-\eta/\nu}\ , \eeq with
parameters $\epsilon=0.361$, $\eta=5$ and $\nu=1.2$. The comoving
cutoff scale $R_c^0$, defined so that at $k=1/R_c^0$ the power
spectrum is reduced in half compared to CDM, is given by
\begin{eqnarray} R_c^0 & = & 0.201 \left( \frac{ \Omega_X h^2} {0.15}
\right)^{0.15} \left( \frac{g_X} {1.5} \right) ^{-0.29} \left(
\frac{m_X} {\rm 1~keV} \right)^{-1.15}~{\rm Mpc}\ \nonumber \\ & = &
0.226 \left( \frac{ \Omega_X h^2} {0.15} \right)^{-0.14} \left(
\frac{\vzrms} {\rm 0.05~km/s} \right)^{0.86}~{\rm Mpc}\
. \label{eq:Rc} \end{eqnarray} In our calculations below we also
require the WDM power spectrum at matter-radiation equality, $z=z_{\rm
eq}$. Based on results from a Boltzmann code (C.-P.\ Ma, personal
communication), we find for $\Omega_X=0.3$ and $h=0.7$ an accurate
fitting formula for $T_X(k)$ of the same form as above, but with
parameters $\epsilon=0.359$, $\eta=3.81$ and $\nu=1.1$. In this case,
the cutoff scale is $R_c^{\rm eq} = 0.932 R_c^0$. As expected, at
equality the cutoff in the power spectrum is less sharp than at $z=0$,
and so $\eta$ is smaller. At either redshift, we can define a
characteristic mass $M_c$ from the characteristic scale $R_c$ using
the mean density of the universe, and thus \beq M_c = 1.74 \times 10^8
\left( \frac{ \Omega_0 h^2} {0.15} \right) \left( \frac{ R_c} {\rm
0.1~ Mpc} \right)^3~M_{\sun}\ . \eeq

As we discuss in detail below, the effect of free streaming is to
suppress the formation of small WDM halos. This is especially true at
high redshift, when only the rare, most overdense peaks collapse, and
even a small delay in the collapse of each halo can greatly reduce the
abundance of halos. Since these small halos merge to create the dense
cores of later halos in CDM models, low-redshift halos in WDM are less
centrally-concentrated and have shallower density profiles. Numerical
simulations confirm that WDM models provide a closer match to
observations of dwarf galaxies, in terms of their abundance and
structure \citep{colin00, Bode, eke01}, the sizes of their gaseous
disks \citep{larsen}, and their spatial distribution and formation
epoch \citep{Bode}. The observed scaling of central phase-space
density with mass, in objects ranging from dwarf galaxies to galaxy
clusters, is inconsistent with the simple idea of an initial
phase-space density which remains conserved; however, only the
averaged, coarse-grained phase-space density is accessible to
observations, which therefore imply that the coarse-grained density of
most objects must have significantly decreased during the hierarchical
formation process \citep{dalcanton, sellwood}. Still, since the
phase-space density can only decrease (in the absence of dissipation),
the high phase-space densities of dwarf spheroidal galaxies imply a
minimum mass of $m_X \sim 0.7$ keV for thermal WDM \citep{dalcanton}.
The precise limit depends on the assumption that the DM core radius in
the dwarf spheroidals is roughly twice the observed core radius in the
stellar distribution. If the ratio is four rather than two then the
mass limit weakens to $m_X \sim 0.4$ keV. An independent limit on
$m_X$ has been derived based on a comparison of $z \sim 3$
observations and numerical simulations; requiring the WDM model to
match the observed opacity distribution of the Lyman-$\alpha$ forest
results in a lower limit of $m_X \sim 0.75$ keV \citep{vijay00}.

In this paper we derive a new, independent constraint on WDM models.
If the problem of dwarf galaxy density profiles is solved by invoking
WDM, the resulting strong suppression of the abundance of dwarf
galaxies at high redshift may be inconsistent with the observed fact
of the reionization of the universe, since precisely these galaxies
must produce the reionization.

\section{Modeling Methods}

\subsection{Halo Formation in WDM}

\label{sec:Formation}

To understand the properties of halos in WDM, the non-linear problem
of gravitational collapse must be considered. The dynamical collapse
of a dark matter halo can be solved analytically only in cases of
particular symmetry. In particular, spherical collapse has proven very
useful in understanding the properties and distribution of halos in
CDM, particularly since the results enter into the Press-Schechter
model for the halo abundance. Thus, as a first step in building a
semi-analytic model for WDM, we consider in this section the formation
of WDM halos under spherical symmetry.

In CDM, spherical collapse is simple enough that it can be solved
analytically. For an initial top-hat perturbation, different
mass-shells do not cross until the final collapse and
virialization. The motion of each shell is simply determined by the
enclosed mass, which is fixed, and thus the turnaround and collapse
times can be calculated, with the final radius derived from the radius
at turnaround together with the virialization condition. The
corresponding problem in WDM is, however, much more challenging, even
once a numerical approach is adopted. The particles making up each
mass shell start out with random velocities as given by the
Fermi-Dirac distribution function (\S \ref{sec:WDM}). Even in the
initial stages, particles from different mass shells mix, and the
velocity distribution function at each radius changes shape. For an
accurate numerical solution, the full velocity distribution function
at each radius must be well resolved, as opposed to the single shell
velocity which suffices in the CDM case. We defer to later work the
full implementation of WDM in spherical collapse simulations, and
instead we focus on a simpler model which closely approximates the
physics of WDM.

This alternative physical system, which we use as a model for WDM, is
an adiabatic gas, with an initial temperature which corresponds to the
initial velocity dispersion of the WDM. We emphasize for clarity that
this gas does not correspond to baryons but is instead a model for
WDM. To set up the analogy in the initial nearly-homogeneous universe,
we match the root-mean-square velocities of the (non-relativistic) WDM
and the analogous gas. This requires a gas temperature $T$ where \beq
\vrms^2=3 k_B T/ \mu\ , \eeq in terms of the mean molecular mass $\mu$
chosen for the gas. The conservation of phase-space density for WDM
yields $\vrms \propto \rho^{1/3}$, equivalent to an adiabatic gas with
pressure $p \propto \rho^{\gamma}$ where the index $\gamma=5/3$. It is
easy to show that, in the simple case of a homogeneous expanding
universe, the WDM and gas models are mathematically identical. The
decline of $\vrms$ as $1/a$ for WDM corresponds to the adiabatic
cooling $T \propto 1/a^2$ for the gas. During the collapse, the
conservation of phase-space density yields an effective equation of
state, although the analogy to gas is not exact because it does not
precisely capture the effect of shell mixing and it neglects
anisotropic stress. Furthermore, there is a fundamental difference in
the behavior of small-scale modes in the two cases; these modes decay
in the case of free streaming, but they remain constant in amplitude
and oscillate as sound waves in the case of gas pressure. Nonetheless,
for our purposes the key is in the similarity of free streaming and
gas pressure when they compete with gravity; both effects allow the
growth of fluctuations above a characteristic mass scale and prevent
it below that scale. Numerically, the required initial gas temperature
at $z=z_i$ is \beq T=9.1\times 10^5 \left(\frac{\mu} {m_p}\right)
\left(\frac{\vzrms} {\rm 0.05~km/s}\right)^2 \left(\frac{1+z_i} 
{3000} \right)^2 \, K\ , \eeq where $m_p$ is the proton mass.

We simulate spherical WDM/gas collapse using an improved version of
the code presented in \citet{htl} and originally developed by
\citet{tw95}. This is a one-dimensional, spherically symmetric
Lagrangian hydrodynamics code. For initial conditions at time $t_i$ we
assume spatial fluctuations in the WDM density in the form of a single
spherical Fourier mode, \beq \frac{\rho(r,t_i)-\bar{\rho}(t_i)}
{\bar{\rho} (t_i)} = \delta(t_i)~ \frac{\sin (kr)}{kr}\ , \eeq where
$\bar{\rho}(t_i)$ is the background density of WDM at $t_i$,
$\delta(t_i)$ is the initial overdensity amplitude, $r$ is the
comoving radial coordinate, and $k$ is the comoving perturbation
wavenumber. We define the halo mass as the initial mass inside the
radius given by $k r=\pi$. For the initial redshift $z_i$ we choose
matter-radiation equality, \beq 1+z_{\rm eq}=3600 \left( \frac{\Omega
h^2} {0.15} \right)\ . \eeq Since for reionization we must be able to
predict the halo mass function as early as $z=20$--30, in order to
begin in the linear regime we must choose $z_i \gg 100$. Furthermore,
as shown below, most of the effect of WDM that causes the delay in
halo collapse occurs at very high redshift, so in order to include the
entire effect it is necessary to begin at $z_{\rm eq}$ at which time
density perturbations begin to grow significantly.

\begin{figure}
\plotone{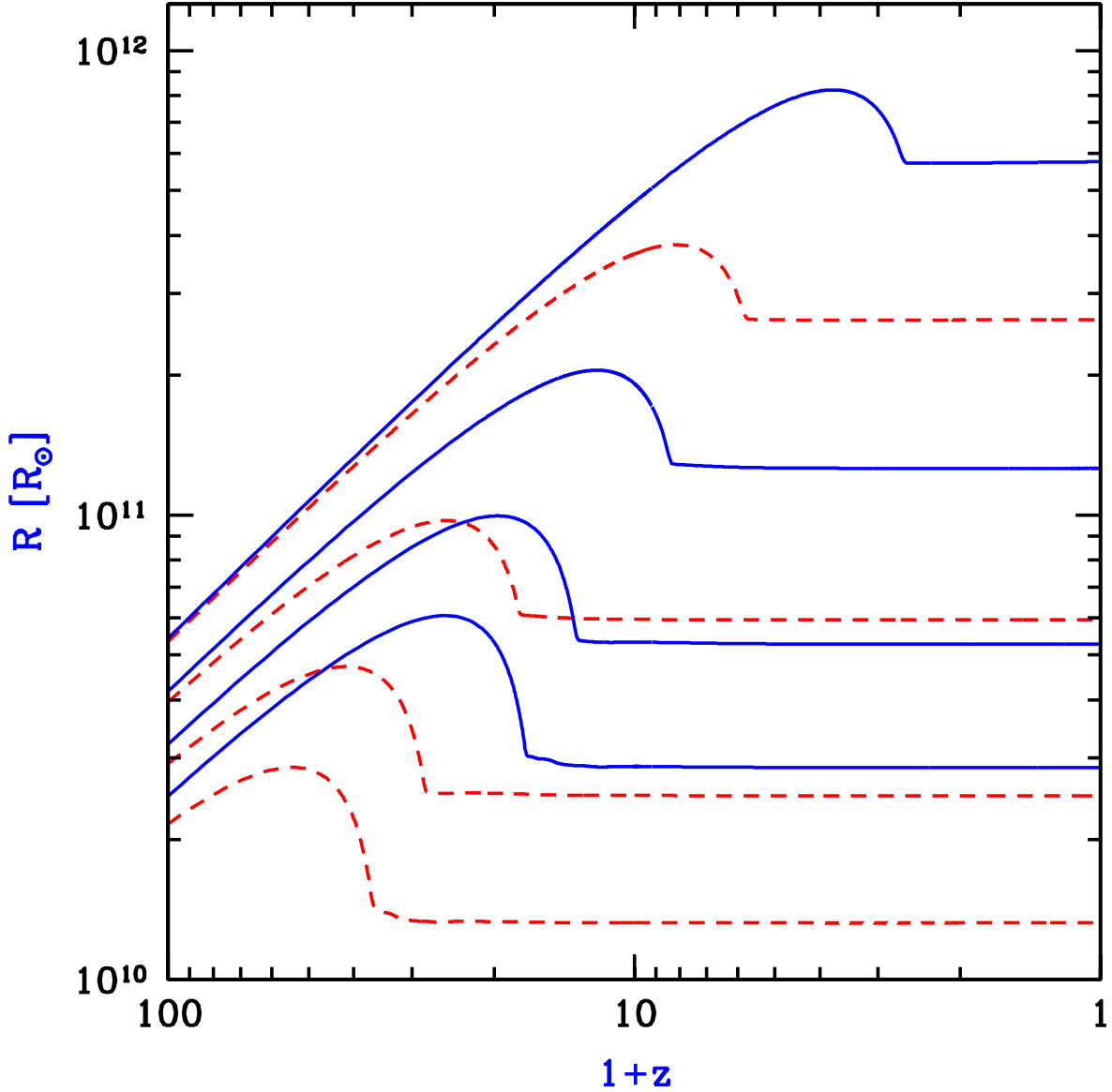} 
\caption{Example of shell trajectories for gas with initial
temperature corresponding to $\Lambda$WDM with $m_X=1$ keV
($\vzrms=0.041~$km/s; solid curves) and for gas with initial $T=0$ K
(corresponding to $\Lambda$CDM; dashed curves). In each case, four
trajectories are shown, in order of increasing enclosed mass (from
bottom to top). The curves correspond to enclosed masses of $1/4$,
$1/2$, 1 and 2, respectively, in units of $1.6\times 10^8 M_{\odot}$
which is the mass of the halo in this case.
\label{fig:traj}}
\end{figure}

Figure~\ref{fig:traj} shows an example of shell trajectories for gas
corresponding to $\Lambda$WDM with $m_X=1$ keV ($\vzrms=0.041~$km/s;
solid curves) and for gas corresponding to $\Lambda$CDM (dashed
curves). The trajectories are shown for four different mass shells,
where the second-highest shell corresponds to the halo mass of
$1.6\times 10^8 M_{\odot}$. Each shell reaches a radius of maximum
expansion and then contracts until it is stopped by the outward-moving
virialization shock. In the case of WDM particles a similar trajectory
shape would arise, if we consider the radius of a shell enclosing a
fixed mass, with shocking replaced by violent relaxation occurring due
to rapid shell mixing. By comparing the different curves in the
figure, we calculate the delay relative to $\Lambda$CDM in the
virialization redshift, caused by the WDM velocities. For the
different mass shells shown in the figure, the delay factor in the
value of $(1+z)$ at virialization ranges from 1.99--2.09, and the
increase in the radius at virialization ranges from 2.12--2.19. Each
of these ranges is narrow, and thus the results are insensitive to the
halo density profile. Also, the two factors (for $(1+z)$ and for the
radius) are nearly equal, which implies that the virialization radius
in WDM is smaller than it would be in CDM, for the {\it same}\,
virialization redshift, by only $\sim 5\%$ for this halo mass. We find
in general that although the halo virialization is delayed in WDM, and
the final halo radius is larger than in CDM, the homogeneous
background universe also expands significantly during this extra
period; the final halo overdensity relative to the cosmological mean
at the virialization time is almost unchanged, and we neglect the
small change that may occur.

In this gas model of WDM, the delayed collapse of halos is due to
pressure, and the characteristic halo mass below which a significant
effect occurs is the Jeans mass. In general, the Jeans mass $M_J$ is
the halo mass for which pressure just balances gravity initially (in
the linear regime). For a temperature $T$ and density $\rho$, $M_J
\propto T^{3/2}/\rho^{1/2}$. For WDM this yields, \begin{eqnarray} M_J
& = & 3.06 \times 10^8 \left( \frac{g_X} {1.5} \right)^{-1} \left(
\frac{ \Omega_X h^2} {0.15} \right)^{1/2} \left( \frac{m_X} {\rm
1~keV} \right)^{-4} \left( \frac{1+z_i}{3000} \right)^{3/2}~M_{\sun}\
\nonumber \\ & = & 4.58 \times 10^8 \left( \frac{ \Omega_X h^2} {0.15}
\right)^{-1/2} \left( \frac{\vzrms} {\rm 0.05~km/s} \right)^{3} \left(
\frac{1+z_i}{3000} \right)^{3/2}~M_{\sun}\ . \label{MJeans}
\end{eqnarray} The collapse simulations confirm these scalings. 
Indeed, for a given collapse delay factor, the affected halo mass
varies with $m_X$ and with $z_i$ just as $M_J$ does. In particular, it
is necessary to start the simulations at $z_{\rm eq}$, since the
effective pressure of WDM affects the highest halo masses at the
highest redshifts. If we were to begin the simulations at a lower
$z_i$, with identical initial conditions in WDM and CDM, the
comparison would not include the full effect of WDM, since it would
not include the effect at $z>z_i$ of WDM pressure on the collapse. 

\subsection{Halo Abundances in WDM}

\label{sec:Abundance}

In addition to characterizing the properties of individual halos, a critical
prediction of any theory of structure formation is their mass function, i.e.
the abundance of halos as a function of mass, at each redshift. This prediction
is an important step towards inferring the abundance of galaxies. While the
number density of halos can be measured for particular cosmologies in numerical
simulations, an analytic model helps us gain physical understanding and can be
used to explore the dependence of abundances on redshift and on all model
parameters.

A simple analytic model which successfully matches many numerical
simulations of CDM was developed by \citet{PS}. The model is based on
the ideas of a Gaussian random field of initial density perturbations,
linear gravitational growth, and spherical collapse. The abundance of
halos at redshift $z$ is determined using $\delta_M$, the density
field smoothed on a mass scale $M$.  The variance $\langle \delta_M
\rangle^2$ is denoted $\sigma^2(M)$, where the conventional filter is
a top-hat of comoving radius $R$, with $M=4 \pi \rho_m R^3/3$, in
terms of the current mean density of matter $\rho_m$. Although the
Press-Schechter model is based on the initial conditions, it is
usually expressed in terms of redshift-zero quantities. Thus, it uses
the linearly-extrapolated density field, i.e., the initial density
field at high redshift extrapolated to the present by simple
multiplication by the relative growth factor.

Since $\delta_M$ is distributed as a Gaussian variable with zero mean
and standard deviation $\sigma(M)$ [which depends only on the present
linearly-extrapolated power spectrum], the probability that $\delta_M$
is greater than some threshold $\delta$ equals \beq
\int_{\delta}^{\infty} d\, \delta_M \frac{1}{\sqrt{2 \pi}\, \sigma(M)}
\exp \left[- \frac{\delta_M^2} {2 \,\sigma^2(M)}\right]={1\over 2} \,
{\rm erfc}\left[\frac{\delta} {\sqrt{2} \,\sigma(M) } \right]\
. \label{PS1} \eeq The fundamental ansatz of Press \& Schechter was to
identify this probability with the fraction of DM particles
which are part of collapsed halos of mass greater than $M$ at redshift
$z$, with two additional ingredients: First, the value used for
$\delta$ is $\delta_{\rm crit}(z)$, which is the critical density of
collapse found for a spherical top-hat (extrapolated to the present
since $\sigma(M)$ is also calculated using the linear power spectrum
extrapolated to the present); and second, the fraction of DM
in halos above $M$ is multiplied by an additional factor of 2 in order
to ensure that every particle ends up as part of some halo with
$M>0$. Thus, the final formula for the mass fraction in halos above
$M$ at redshift $z$ is \beq \label{PSerfc} F(>M | z)={\rm
erfc}\left(\frac{\nu} {\sqrt{2}} \right)\ , \eeq where
$\nu=\delta_{\rm crit}(z)/ \sigma(M)$ is the number of standard
deviations which the critical collapse overdensity represents on mass
scale $M$. Expressing this as $F(> \nu | z)$ yields the mass function
\beq \nu f(\nu)\equiv -\nu \frac{dF(> \nu | z)}{d\nu}=\sqrt{\frac{2}
{\pi}}\, \nu\, \exp \left[ -\frac{\nu^2}{2} \right]\ . \eeq The
comoving number density $dn$ of halos of mass between $M$ and $M+dM$
is given by \beq\frac{dn}{dM}= \frac{\rho_m}{M}\, \frac{-d(\ln
\sigma)}{dM} \nu f(\nu)\ . \eeq

As noted above, the ad-hoc correction by a factor of two is necessary,
since otherwise only positive fluctuations of $\delta_M$ would be
included. \citet{Bond} found an alternate derivation of this
correction factor, using a different ansatz. In their derivation, the
factor of two has a more satisfactory origin, namely the so-called
``cloud-in-cloud'' problem: For a given mass $M$, even if $\delta_M$
is smaller than $\delta_{\rm crit}(z)$, it is possible that the
corresponding region lies inside a region of some larger mass $M_L>M$,
with $\delta_{M_L}>\delta_{\rm crit}(z)$. In this case the original
region should be counted as belonging to a halo of mass $M_L$. Thus,
the fraction of particles which are part of collapsed halos of mass
greater than $M$ is larger than the expression given in
equation~(\ref{PS1}).

\citet{Bond} considered how to determine the mass of the halo which
contains a given point $\bf{O}$. They examined the value of $\delta_M$
as a function of $M$ about the point $\bf{O}$, starting with
$\delta_M=0$ as $M \rightarrow \infty$, with $\delta_M$ increasing as
smaller regions are considered. They proposed to find the largest
value of $M$ for which $\delta_M$ is sufficiently large [i.e., above
$\delta_{\rm crit}(z)$] to correspond to a collapsed halo. Since this
mass corresponds to the largest collapsed halo around $\bf{O}$,
associating this halo with the point $\bf{O}$ naturally solves the
cloud-in-cloud problem. To further simplify the derivation, in the
smoothing kernel used to define $\delta_M$ \citet{Bond} used a
$k$-space top-hat filter rather than a top-hat in real space. This
way, as $M$ is lowered, new $k$ values enter which, for a Gaussian
random field, are uncorrelated with those previously included. This
results in a random walk whose statistics are determined by the
variance $S=\sigma^2(M)$.  Starting at $\delta_M=0$ when $S=0$, the
first crossing of $\delta_M$ above the barrier $B=\delta_{\rm
crit}(z)$ yields the mass of the halo containing $\bf{O}$. Performing
many such random walks results in the halo abundance, where the mass
fraction in halos with mass between $M_1$ and $M_2$ equals the
fraction of random walks which first cross above the barrier $B$
between $S(M_1)$ and $S(M_2)$. Since the barrier is constant (i.e.,
independent of $M$) in CDM, \citet{Bond} could derive the halo
abundance analytically, and showed that this solution to the
cloud-in-cloud problem results precisely in a factor-of-two correction
to equation~(\ref{PS1}). Although the \citet{Bond} derivation uses
$\sigma(M)$ as defined by a $k$-space top-hat filter, the final
formula is usually applied using a real-space top-hat. We follow this
convention throughout this paper.

The cloud-in-cloud correction is more complicated in the case of WDM.
As shown in the previous subsection, halo collapse is delayed when the
velocity dispersion of WDM is included. Thus, the collapse threshold
$\delta_{\rm crit}$ becomes a function of $M$ as well as $z$ (Note
that the linear extrapolations used to define $\delta_{\rm crit}$ and
$\sigma(M)$ use the growth factor as calculated in $\Lambda$CDM).  A
grid of collapse simulations in ($z,M$) as described above, can be
used to derive the full function $\delta_{\rm crit}(z,M)$ for
different values of $m_X$. In addition, the cut-off in the power
spectrum due to free streaming by the WDM particles lowers the value
of $\sigma(M)$ compared to that in CDM. Note that since we determine
$\delta_{\rm crit}$ based on simulations that are run from
matter-radiation equality, for consistency we calculate $\sigma(M)$
based on the WDM power spectrum at equality. In WDM, $\sigma(M)$
approaches a constant value as $M \rightarrow 0$, unlike CDM where it
continues to rise logarithmically. On the other hand, the barrier
height $\delta_{\rm crit}$ diverges in WDM as $M \rightarrow 0$, while
it is constant in CDM. Thus, in CDM the expression on the right-hand
side of equation~(\ref{PS1}) increases monotonically as the mass is
lowered, while in WDM this same expression (with $\delta=\delta_{\rm
crit}$) reaches a maximum and then declines toward zero as $M
\rightarrow 0$. Thus, a naive application to WDM of
equation~(\ref{PS1}) fails, and the cloud-in-cloud correction is
crucial in this case. We derive the halo mass fraction $F(>M | z)$ in
WDM by numerically generating random walks and counting the
distribution of their first crossings of the barrier $B=\delta_{\rm
crit}(z,M)$.

An additional ingredient is necessary for calculating an accurate mass
function. Even in $\Lambda$CDM, the Press-Schechter halo mass function
disagrees somewhat with that measured in numerical simulations.
Specifically, the simulations find larger numbers of rare, massive
halos but smaller numbers of the more abundant low-mass halos.
\citet{st98} fitted the mass function seen in simulations with a
function of the form: \beq \nu f(\nu)=\sqrt{2}{\pi} A \left(1+
\frac{1} {\hat{\nu}^{2 q}} \right) \hat{\nu} \exp\left( -\frac{
\hat{\nu}^2} {2}\right)\ , \label{sheth}\eeq where
$\hat{\nu}=\sqrt{a}\nu$, and the fitted parameters are $a=0.707$,
$q=0.3$, and $A=0.322$.  \citet*{smt00} showed numerically that this
correction to the Press-Schechter mass function is equivalent to
changing the barrier shape from the case of the constant barrier
$B_0=\delta_{\rm crit}(z)$. Namely, the ``naive'' barrier $B_0$ must
be mapped to the actual barrier by \beq B=\sqrt{a}\, B_0 \left[1+b
\left( \frac{\sigma^2(M)} {a B_0^2} \right)^c\, \right] \ ,
\label{barrier} \eeq where $a$ is the same as before and also $b=0.5$
and $c=0.6$. \citet{smt00} also show that if the effect of shear and
ellipticity on halo collapse is considered, the resulting barrier $B$
is modified from $B_0$ (which is based on spherical collapse) in a way
that is similar to the above fit, except with $a=1$. In particular,
ellipticity does not explain the under-prediction of the number of
massive halos. Thus, ellipticity explains the need for a more
complicated barrier shape than is suggested by spherical halo
collapse, but the precise shape needed to fit the simulations may
indicate the presence of additional effects.

\begin{figure} 
\plotone{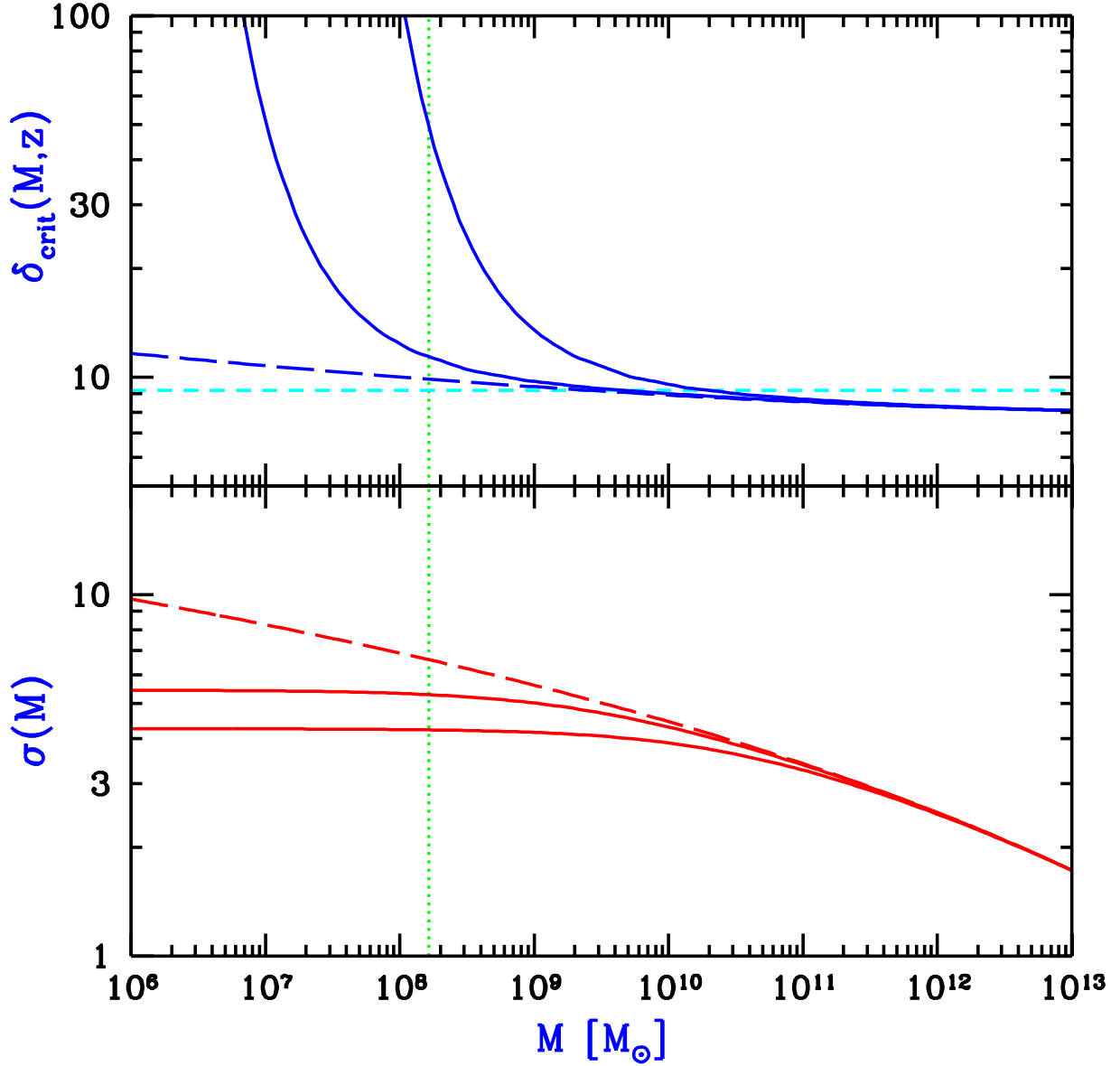} 
\caption{Halo formation in WDM, at $z=6$. In the top panel, we show
the linear extrapolated overdensity $\delta_c(M,z)$ at the time of
collapse, as a function of halo mass $M$. The solid curves show the
cases of $m_X=1.5$ keV and $m_X=0.75$ keV, respectively from bottom to
top. For comparison, we show the $\Lambda$CDM curve which includes the
Sheth/Jenkins correction (long-dashed curve), and the mass-independent
value given by spherical collapse in $\Lambda$CDM (short-dashed
curve). In the bottom panel, we show the mass fluctuation $\sigma(M)$,
based on the linearly-extrapolated power spectrum at matter-radiation
equality. The solid curves illustrate the effect of the power spectrum
cutoff in $\Lambda$WDM, for $m_X=1.5$ keV, and $m_X=0.75$ eV,
respectively from top to bottom. Also shown for comparison is
$\sigma(M)$ in $\Lambda$CDM (long-dashed curve). In both panels, the
vertical dotted line shows the value of the lowest halo mass at $z=6$
in which gas can cool (see \S \ref{sec:reion}).
\label{fig:deltac}}
\end{figure}

Regardless of the precise physical explanation, the mapping given by
equation~(\ref{barrier}) can be applied to produce mass functions
which agree with numerical simulations of $\Lambda$CDM. Assuming that
the same mapping applies to WDM as well, we apply this mapping to the
``naive'' WDM barrier $B_0=\delta_{\rm crit}(z,M)$ to derive the final
barrier shape. More precisely, we use the mass function with slightly
different parameters as specified by \citet{jenk00}, who fitted a
number of simulations with a larger range of halo mass and redshift.
We note that \citet{jenk00} adopted a somewhat unconventional
definition of halo mass when they derived their mass function
(enclosing 180 times the background density, as in the
Einstein-deSitter model, rather than the value suggested by spherical
collapse in $\Lambda$CDM). In the present application at high
redshift, where $\Omega(z)\approx 1$, this has little effect on the
mass function \citep[see also][for a general discussion of the effect
of the halo mass definition on the mass function]{white00}. The
fitting parameters found by \citet{jenk00} are $a=0.73$, $q=0.175$,
and $A=0.353$ in equation~(\ref{sheth}). We find numerically that this
mass function is approximately generated if $b=0.34$ and $c=0.81$ are
chosen in equation~(\ref{barrier}).  Figure~\ref{fig:deltac} shows the
resulting value of $\delta_{\rm crit}$ (top panel) as a function of
halo mass $M$, at $z=6$. Also shown (bottom panel) is the mass
fluctuation $\sigma(M)$. The $\Lambda$WDM models correspond to
$m_X=1.5$ keV ($\vzrms=0.024~$km/s; $R_c=0.12\,$Mpc; $M_c=3.1 \times
10^8 M_{\odot}$) and $m_X=0.75$ keV ($\vzrms=0.060~$km/s;
$R_c=0.27\,$Mpc; $M_c=3.4 \times 10^9 M_{\odot}$), where the $z=0$
power spectrum has been used to define $R_c$ and $M_c$. In the
$\Lambda$WDM models, the figure shows a rapid rise in $\delta_c$ along
with a saturation of $\sigma(M)$ below the cutoff mass. In both
panels, the vertical dotted line shows the value of the lowest halo
mass at $z=6$ in which gas can cool (see \S \ref{sec:reion}).

\begin{figure} 
\plotone{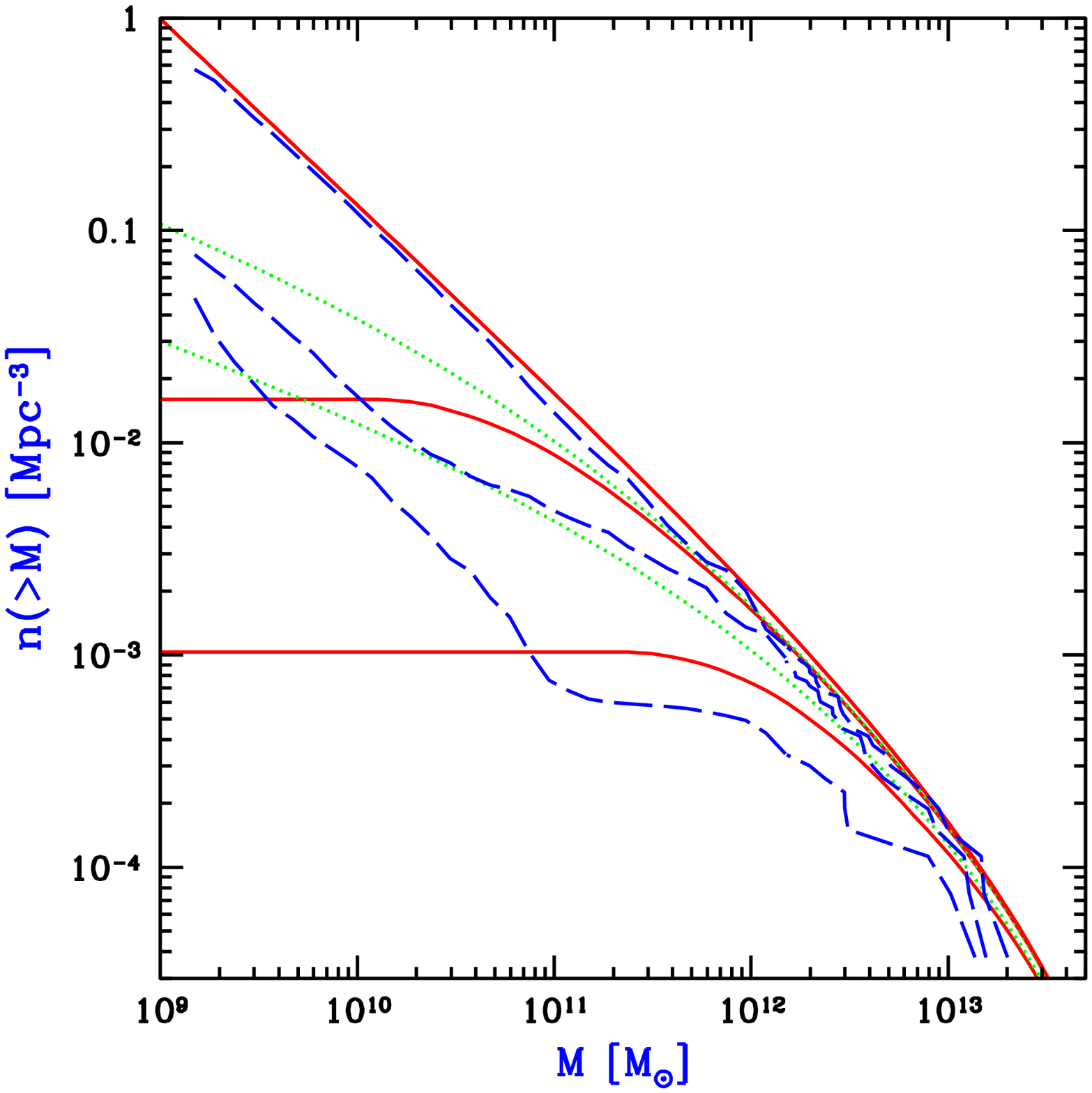} 
\caption{Numerical and semi-analytic halo mass functions at $z=1$ in
$\Lambda$CDM and in $\Lambda$WDM. The curves show the comoving number
density $n(>M)$ of halos above mass $M$. Solid lines show the
semi-analytic calculation, and dashed lines show the results from
numerical simulations by \citet{Bode}. The three cases shown are, from
top to bottom, $\Lambda$CDM, $m_X=0.35$ keV, and $m_X=0.175$ keV. The
dotted lines show model predictions, for the two $\Lambda$WDM models,
which include only the suppression of the power spectrum. The
lowest-mass halos, those below $\sim 10^{10} M_{\odot}$ (for
$m_X=0.35$ keV) or $\sim 10^{11} M_{\odot}$ (for $m_X=0.175$ keV),
form in the simulations through fragmentation and are not accounted
for in our models, but these halos do not affect our conclusions about
reionization (see text). Note that the models shown in this figure use
the cosmological parameters $\Omega_{\rm b}=0$ and $h=0.67$ for
consistency with \citet{Bode}.
\label{fig:Bode}}
\end{figure}

Like the other assumptions which enter into our semi-analytic model,
our use of equation~(\ref{barrier}) is ultimately justifiable by
comparing the final result to numerical simulations. Recently,
\citet{Bode} performed simulations down to $z=1$ for $\Lambda$WDM with
$m_X=0.35$ keV and $m_X=0.175$ keV, and compared them with a
simulation of $\Lambda$CDM. In their Figure~9, \citet{Bode} show the
halo mass function, in the form of the comoving number density $n(>M)$
of halos above mass $M$. We compare their results to our models in
Figure~\ref{fig:Bode}. The model predictions which include only the
suppression of the power spectrum (dotted lines) clearly do not
produce enough suppression to match the simulations. The full model,
which also accounts for the delay of halo collapse, still predicts
less suppression than is seen in the simulations but matches more
closely the simulated mass function over a range of scales for which
the halo abundance is significantly suppressed compared to
$\Lambda$CDM. The lowest-mass halos disappear in the $\Lambda$WDM
models, while the simulations produce significant numbers of halos
below $\sim 10^{10} M_{\odot}$ (for $m_X=0.35$ keV) and $\sim 10^{11}
M_{\odot}$ (for $m_X=0.175$ keV). As noted by \citet{Bode}, these
halos do not form hierarchically, and thus are not accounted for in
our models. These halos do not, however, affect our conclusions below
about reionization. Such halos form by top-down ``pancake'' (actually
ribbon) fragmentation at low redshift, and are not present in
significant numbers at the high redshifts which are relevant for the
reionization of the universe. In particular, the rare high-$\sigma$
peaks around which galaxies form at high redshift tend to be more
nearly spherical and more centrally concentrated than low-$\sigma$
peaks \citep{BBKS}. This should tend to reduce the importance of
fragmentation at high redshift. 

\begin{figure} 
\plotone{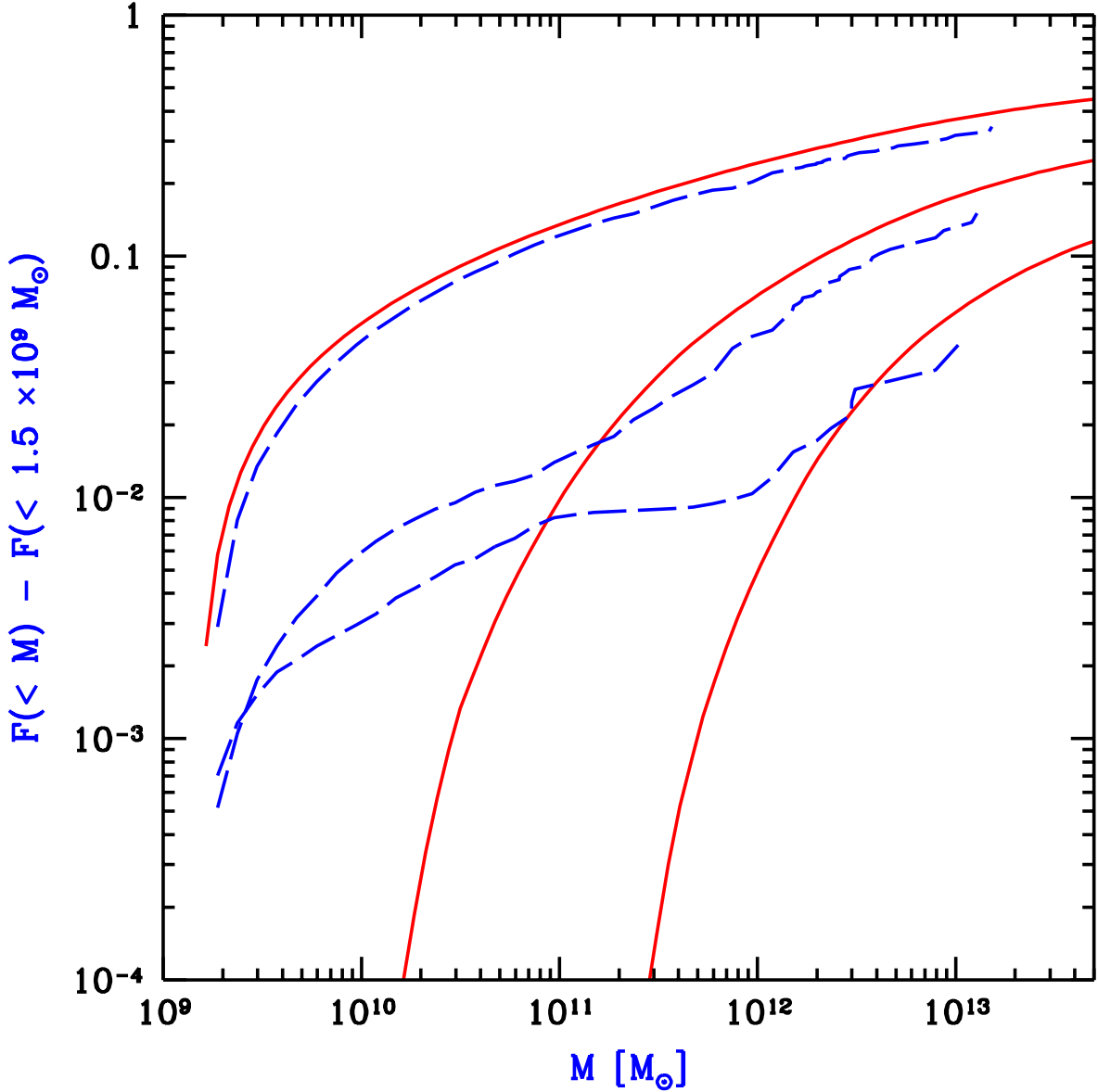} 
\caption{Numerical and semi-analytic halo mass fractions at $z=1$ in
$\Lambda$CDM and in $\Lambda$WDM. The curves show the total mass
fraction $F(<M)$ in halos up to mass $M$, with $F(<1.5 \times 10^9
M_{\sun})$ subtracted. Solid lines show the semi-analytic calculation,
and dashed lines show the results from numerical simulations by
\citet{Bode}. The three cases shown are, from top to bottom,
$\Lambda$CDM, $m_X=0.35$ keV, and $m_X=0.175$ keV. Note that the
models shown in this figure use the cosmological parameters
$\Omega_{\rm b}=0$ and $h=0.67$ for consistency with \citet{Bode}.
\label{fig:BodeNew}}
\end{figure}

Even at low redshift, though, our models correctly predict the total
mass fraction in halos, which is the quantity that is directly
relevant to reionization. This successful prediction is shown in
Figure~\ref{fig:BodeNew}, which presents differently the same data as
in Figure~\ref{fig:Bode}. The quantity shown is the total mass
fraction in halos up to halo mass $M$, as a function of $M$, except
that we subtract out the mass fraction in halos below a mass of $1.5
\times 10^9 M_{\sun}$, which is roughly the minimum mass of halos
which are well resolved in the simulations. As before, the good
agreement is apparent for the $\Lambda$CDM model. In the WDM cases,
the semi-analytic models give a much lower mass fraction than the
simulations, in halos at the low-mass range, but when the mass is
accumulated up to the highest-mass halos, the total mass fraction is
in fact slightly higher in the models than in the simulations. Thus,
in the simulations there is a shifting of mass from large halos to
small halos, in that some regions that should collapse whole,
according to the semi-analytic model, actually fragment and collapse
as several sub-halos. The overall mass fraction in halos, however, is
rather accurately reproduced by the model.

We emphasize that the comparison of the semi-analytic mass function to
the one derived from simulations must be considered preliminary. In
the model, we have introduced a number of approximations, while the
WDM simulations must also resolve a number of issues. For example,
since some halos form through fragmentation, a clear comparison must
be made of the results from different possible ways of identifying
virialized halos in the simulations. In addition, sufficiently small
scales are already weakly nonlinear at $z \sim 40$, and the neglect of
non-linear effects can only be checked by making runs which start at a
higher redshift. 

As shown in \S \ref{sec:Formation}, the collapse of a given halo must
be followed from a very high redshift, near matter-radiation equality,
in order for the full delay caused by the WDM velocities to be
included. However, even though the numerical simulations are begun at
$z \sim 40$, they do in fact include the effect of WDM velocities at
higher redshift; the simulations use the appropriate initial power
spectrum at the starting redshift, which implies, by the continuity
equation, the correct initial mean velocity field (which gives the
mean fluid velocity at every point). Therefore, the reduced power in
the initial conditions also implies reduced infall velocities (through
the Z'eldovich approximation), accounting for the cumulative dynamical
effect of the WDM velocities at higher redshift. In our models, if we
were only interested in the halo mass function at $z=1$, we could also
start with the initial power spectrum at $z=40$, and use the ordinary
$\delta_{\rm crit}(z)$, since any remaining WDM velocity dispersion
would only affect extremely small mass scales [note the strong
$z_i$-dependence of the effective Jeans mass, Eq.~(\ref{MJeans})].
Indeed, if random velocities are added to the simulations at $z=40$,
their effect on $z < 40$ dynamics is entirely negligible (Bode 2001,
personal communication). Because the semi-analytic approach treats
changes in $\sigma(M)$ differently from changes in $\delta_{\rm
crit}(z,M)$, the prediction of the halo mass function at $z=1$ would
give somewhat weaker suppression if we started at $z=40$ than if we
start at equality. This points to a certain limitation or ambiguity in
the semi-analytic approach, although we do find that the closest match
to the numerical simulations (in terms of the total mass fraction in
halos) is obtained by starting at equality. In any case, for
reionization we are interested in the halo mass function at $z$ up to
30, and this forces us to choose a very high initial redshift.

\begin{figure} 
\plotone{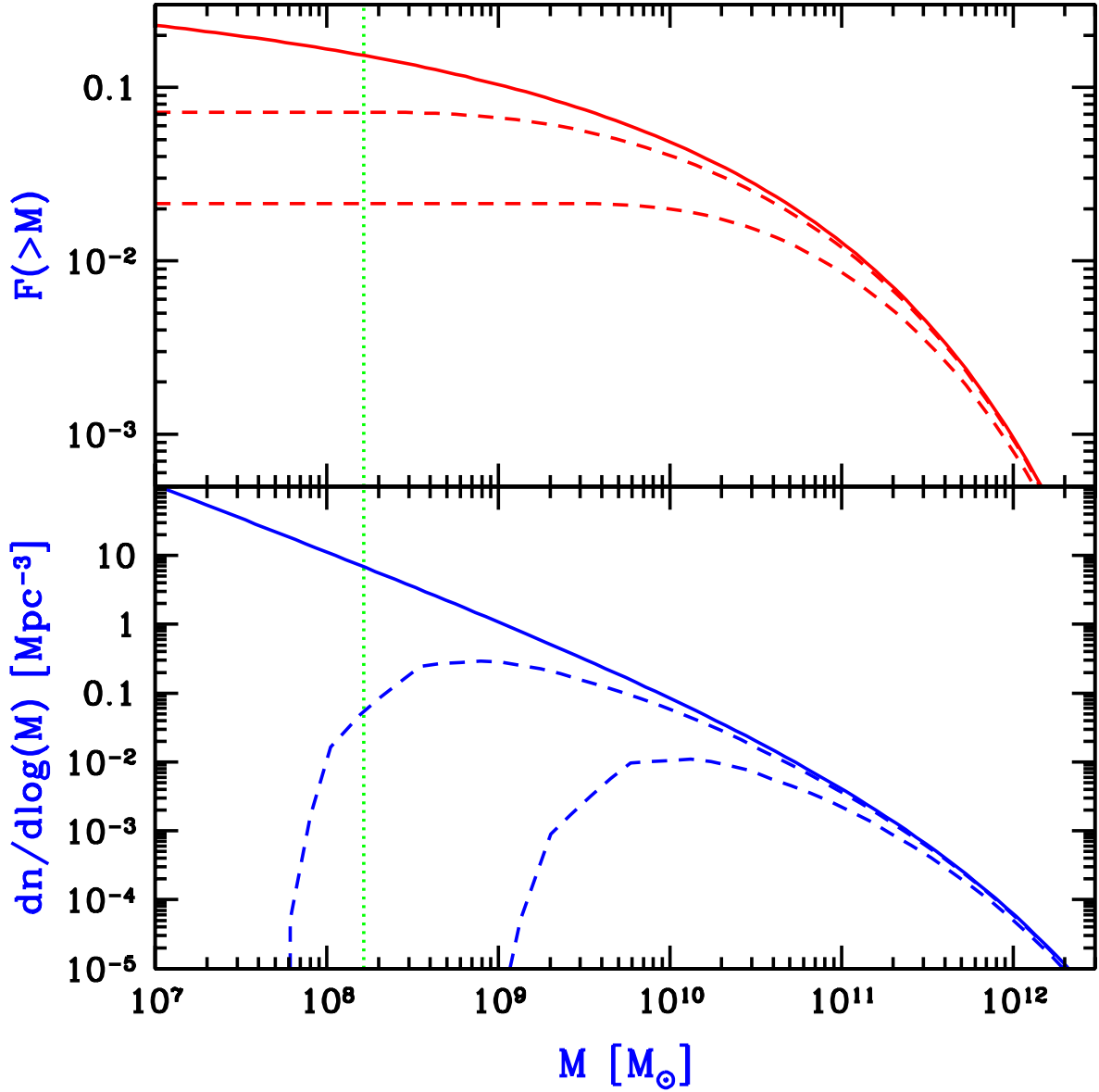} 
\caption{Semi-analytic halo mass functions at $z=6$ in $\Lambda$CDM
and in WDM. The lower panel shows $|dn(>M)/d\log(M)|$, where $n(>M)$
is the comoving number density of halos above mass $M$. The upper
panel shows the corresponding total mass fraction in halos above $M$,
$F(>M)$. In both panels, solid lines are for $\Lambda$CDM and dashed
lines for $\Lambda$WDM with $m_X=1.5$ keV and $m_X=0.75$ keV (top to
bottom). Also shown is the minimum mass for cooling at $z=6$ (vertical
dotted line).
\label{fig:MassFn6}}
\end{figure}

Figure~\ref{fig:MassFn6} shows our model predictions at $z=6$, in
terms of the comoving halo mass function $dn/d\log(M)$ and the total
mass fraction in halos above $M$, $F(>M)$. The figure shows the
comparison for $\Lambda$CDM (solid line) and $\Lambda$WDM (dashed
lines) with $m_X=1.5$ keV and $m_X=0.75$ keV (top to
bottom). Reionization depends on the production rate of ionizing
photons, which in turn depends on the mass fraction in halos in which
gas cools and stars can form (for details see the following
section). The value of the minimum mass for cooling at $z=6$,
corresponding to a virial temperature of $T_{\rm vir}=10^4$K, is also
shown in the figure (vertical dotted line). Clearly, in $\Lambda$WDM
the mass fraction in galaxies is greatly reduced due to a sharp cutoff
in the number density of halos below the cutoff mass. Note that for
$m_X \la 1.5$ keV the cooling mass is unimportant, i.e., even if the
gas could cool in smaller halos, the effect would be negligible
because the abundance of all halos below the cooling mass is severely
reduced (see also Figure~\ref{fig:fcol}).

\subsection{Modeling Reionization}
\label{sec:reion}

We model reionization using a simplified semi-analytic scheme,
following \citet{hl98}. Although this approach does not capture all
the details of the reionization process \citep[see, e.g.,][]{g00}, it
is adequate for our present purpose of estimating the required overall
ionizing-photon budget. We assume that dark halos form at a rate
$d^2n/(dM\,dt)$, where $dn/dM$ is the halo mass function. When a halo
collapses, it converts a fraction of its baryonic mass into either
stars or a central black hole (BH). Each new halo then turns into an
ionizing source, and creates an expanding \HII region. The evolution
of the proper volume $V_i$ of a spherical \HII/\HI ionization front
for a time-dependent source is governed by the equation \beq
\frac{dV_i}{dt}= 3H(z) V_i + \frac{\dot N_{\rm \gamma}(t)}{ \bar
n_{\rm H} } - 1.079\, C\, \bar n_{\rm H}\, \alpha_B V_i,
\label{eq:front} \eeq where $t$ is the age of the source, $H(z)$ is
the Hubble constant, $\bar n_{\rm H}$ is the neutral hydrogen number
density in the smooth IGM at redshift $z$, $\dot N_{\rm \gamma}(t)$ is
the production rate of ionizing photons, $\alpha_{\rm B}=
2.6\times10^{-13}~{\rm cm^3~s^{-1}}$ is the recombination coefficient
of neutral hydrogen to its excited states at $T=10^4$K, and $C\equiv
\langle n_{\rm H}^2 \rangle / \bar n_{\rm H}^2$ is the mean
(volume-weighted) clumping factor of ionized gas within $V_i$ (i.e.,
excluding any dense, self-shielded neutral clumps). In the
recombination term, we have included a factor of 1.079 due to the
additional free electrons produced if helium is singly ionized, as
expected in the case of a reionization dominated by stars. The first
term in equation~(\ref{eq:front}) accounts for the Hubble expansion,
the second term accounts for ionizations by newly produced photons,
and the last term describes recombinations \citep{sg87, hl97}. The
solution of this equation yields the {\it proper}\, ionized volume per
source, and a summation over all sources provides the total ionized
volume within the IGM.

We compute the production rate of ionizing photons for a stellar
population with a \citet{IMF} IMF (initial mass function) undergoing a
burst of star formation at a metallicity equal to $2\%$ of the solar
value \citep{bc96}. We find that the rate is well approximated by \beq
\dot N_{\rm \gamma}(t) = \left\{\matrix{ \dot N_0 \hfill&(t \leq
10^{6.5}~{\rm yr})\hfill\cr \dot N_0 (t/10^{6.5}{\rm yr})^{-4.5}
\hfill&(t > 10^{6.5}~{\rm yr})\hfill,\cr}\right.
\label{eq:rate}
\eeq where $\dot N_0=3.7\times 10^{46}~{\rm s^{-1}~M_\odot^{-1}}$ (per
${\rm M_\odot}$ of stellar mass).  Over the lifetime of the
population, this produces $\approx 4000$ ionizing photons per stellar
proton. We also assume a typical star formation efficiency of $f_*
\equiv M_{\rm stars}/(\Omega_{\rm b}M_{\rm halo}/\Omega_0)=10\%$,
consistent with the observed star formation rate at $z\sim 3$--4
\citep[e.g.,][]{me} and with observations of the metallicity of the
Ly$\alpha$ forest at $z=3$ \citep{hl97}; this efficiency implies that
$\approx 7.2\times 10^{58}$ ionizing photons are produced per ${\rm
M_\odot}$ of total (dark matter + gas) halo mass. We assume further
that the escape fraction of ionizing photons is $f_{\rm esc}=10\%$,
based on observational estimates for present-day galaxies
\citep[e.g.,][but note \citealt*{steidel}]{escape, escape2}. Since the
star formation efficiency and the escape fraction always appear as a
product, we define the parameter $\epsilon_* \equiv f_* f_{\rm esc}$
to parameterize the efficiency of ionizing photon injection into the
IGM. At present, the best direct observational constraint on
$\epsilon_*$ is provided by measurements of the mean ionizing flux at
$z=3$ based on the proximity effect of quasars. We discuss this
further in \S \ref{ReiCons}; we show there that these measurements
imply $\epsilon_*=0.01$, with about a factor of 2 uncertainty.

The analogous $\dot N_{\rm \gamma}(t)$ from gas accretion onto a
quasar BH is not available from an ab-initio theory. However, it can
be derived \citep[for details see][]{hl98} by postulating a BH
formation efficiency ($\epsilon_Q\equiv M_{\rm bh}/M_{\rm halo}$),
adopting an average quasar spectrum \citep[e.g.,][]{elvis94}, and
finding the quasar light-curve that fits the luminosity function of
optical quasars \citep[e.g.,][]{Pei}.  Note that the \citet{elvis94}
spectral template we adopt ($F_\nu \sim \nu^{-1}$) is somewhat harder
than the template ($F_\nu \sim \nu^{-1.8}$) of, e.g., \citet{z98},
which would therefore produce fewer ionizing photons. For a constant
ratio $\epsilon_Q=6\times 10^{-4}$ which is based on observations
\citep{magorrian98}, the above procedure results in $\dot N_{\rm
\gamma}(t)\approx 6.6\times 10^{47} \exp (-t/10^6{\rm yr})~{\rm
s^{-1}}$, yielding approximately $17,000$ ionizing photons per baryon
in the BH; or alternatively $\approx 1.2\times 10^{58}$ per ${\rm
M_\odot}$ of total halo mass. Below we focus on stellar reionization,
and we simply note that quasar BH's would produce $\approx 6$ times
fewer ionizing photons, but their escape fraction may compensate by
being substantially higher than for stars \citep[e.g.,][]{wood00}.

In order to compute the expansion rate of the \HII regions, we must
specify the clumping factor $C$ of ionized gas in
Eq.~(\ref{eq:front}). Here we simply assume a constant value of
$C=10$. For $\Lambda$CDM this is a very conservative value for the
epoch of reionization; since the early ionizing sources form in the
highest-density regions, the clumping factor reaches values as high as
$\approx 100$ in three-dimensional simulations \citep{g00}. Note,
however, that this value for clumping accounts only for absorption by
gas at the highest resolvable density in the simulation. A
higher-resolution simulation would have higher-density gas clumps and
--- depending on the geometry of those clumps --- a higher or possibly
lower clumping factor than the low-resolution simulation. The clumping
factor should also be lower in the case of $\Lambda$WDM, where
clumping in the dark matter is smoothed out on small scales. The
precise effect on the gas clumping factor can only be determined with
numerical simulations or with detailed models.  Below we examine how
our results change when other constant values ($C=1,30$) are assumed.

Finally, to compute the total ionized fraction $F_{\rm H\,II}$, we convolve the
solution of equation~(\ref{eq:front}) with the total cosmological star
formation rate, by 
\beq 
F_{\rm H\,II}(z)= \epsilon_* \int_{0}^{t(z)}dt^{\prime}
\dot \rho_g(t^\prime) {\tilde V}_{\rm H\,II}(t^{\prime},t).
\label{eq:filling}
\eeq Here $t(z)$ is the cosmic time at redshift $z$, $\rho_g(t)=
F_{\rm col} \Omega_{\rm b} \rho_{\rm crit} $ is the total collapsed
gas density (mass per unit comoving volume), $\rho_{\rm crit}$ is the
present-day critical density, and ${\tilde V}_{\rm H\,II}
(t^{\prime},t)$ is the comoving ionized volume per unit stellar mass
at $t$ for a source which emitted at time $t^\prime$. The collapsed
fraction of baryons is evaluated as $F_{\rm col}(z)= F(>M_{\rm
min}(z))$, where the mass fraction $F(>M | z)$ in halos above mass $M$
at redshift $z$ is calculated as in \S \ref{sec:Abundance}. The
minimum mass $M_{\rm min}(z)$ is chosen by requiring efficient
cooling; we assume that the gas within halos smaller than $M_{\rm
min}(z)$ cannot turn into stars. In the metal-deficient primordial
gas, the critical mass $M_{\rm min}(z)$ for cooling corresponds to a
virial temperature of $T_{\rm vir} \approx 10^4\,$K. Unless the
earliest UV sources produce significant X-rays, which then catalyze
the formation of molecular hydrogen, it is unlikely that smaller halos
can contribute to reionization \citep[see][]{har00}. Furthermore, for
the ranges of WDM particle masses we explore here, the abundance of
such small halos is severely suppressed, making their ability to form
stars irrelevant for the constraints we derive (see also
Figures~\ref{fig:MassFn6} and \ref{fig:fcol}).

Based on the spherical collapse simulations (\S \ref{sec:Formation}),
we find that the non-zero velocity of WDM causes a small change in the
$T_{\rm vir}-M_{\rm halo}$ relation.  We find, however, that this
effect is almost negligible. Although the collapse of halos is delayed
relative to their evolution in a CDM universe, once a WDM halo does
collapse, it has nearly the same overdensity relative to the universe
at the collapse redshift as it would have in the CDM case (see the
discussion of Figure~\ref{fig:traj}). Indeed, halos with the same mass
collapsing at the same redshift look nearly indistinguishable in the
WDM and CDM runs; the main difference is that the density fluctuations
producing the WDM halos had a larger initial amplitude. This suggests
that WDM can greatly affect the abundance of halos of a given mass
without significantly affecting their virial overdensity. Note,
however, that in the hierarchical-merging picture WDM may still affect
the inner profiles and the substructure of halos. Material which makes
up the dense core region and some of the substructure within a given
CDM halo originates in small halos at higher redshift, which are later
accreted. If WDM suppresses the abundance of these high-redshift halos
it will affect the inner structure of later halos \citep[see,
e.g.,][]{swamy}.

\section{Results and Discussion}

\subsection{A Constraint from the Quasar Black Hole in SDSS 1044-1215.}

Before describing the constraint on the WDM mass we obtain from
reionization, we first mention a more direct, albeit somewhat weaker
constraint.  The quasar SDSS 1044-1215 discovered by the SDSS at
redshift $z=5.8$ has an apparent $z^\prime$ magnitude of $\approx
19.2$. Under the assumption that this object is neither beamed nor
lensed, and shining at the Eddington luminosity, the mass of the
central BH is inferred to be $M_{\rm bh}=3.4\times 10^9~{\rm M_\odot}$
(see \citealt{f00} and \citealt{hl00} for discussions of how the mass
was derived). If this supermassive BH grew exponentially by accretion
onto a smaller seed BH, the required time for growth is $\ln
(3.4\times 10^9M_\odot/M_{\rm seed})\approx 20$ $e$-folding times for
stellar seeds, where $M_{\rm seed}$ is the mass of the seed. The
natural $e$-folding timescale is the Eddington time, or $4\times 10^7
(\epsilon/0.1)~{\rm yr}$, where $\epsilon$ is the radiative efficiency
of accretion. Assuming $\epsilon=0.1$, and a seed mass of either
$M_{\rm seed}=10~{\rm M_\odot}$ or $M_{\rm seed}=100~{\rm M_\odot}$,
we find that the seed had to be present, and start growing by a
redshift of $z=20$, or $z=15$, respectively.

\begin{figure} 
\plotone{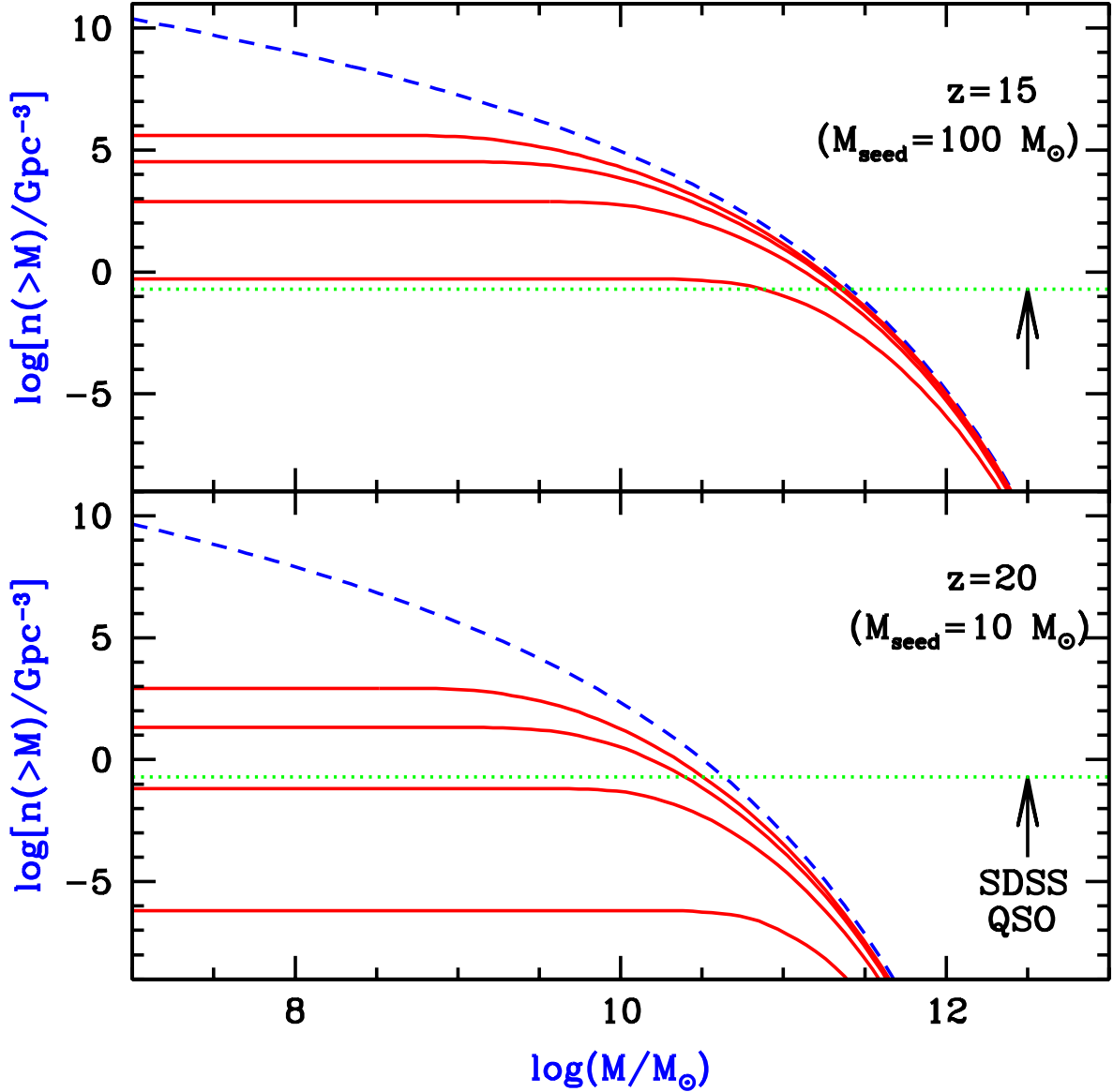} 
\caption{The cumulative mass function of dark halos at two different 
redshifts ($z=15$, upper panel; $z=20$, lower panel).  The dashed
curves show the mass function in the $\Lambda$CDM cosmology, while the
solid curves correspond to mass functions in $\Lambda$WDM models with
$m_X=1.25, 1.00, 0.75$, and $0.50$ keV, respectively from top to
bottom. The horizontal dotted line in each panel corresponds to the
minimum abundance of one halo per 5 Gpc$^3$, required for a host halo
to harbor a seed for the supermassive BH inferred to power the quasar
SDSS 1044-1215.
\label{fig:sdss}}  
\end{figure}

The required presence of a seed by a redshift of $z=20$ or $z=15$
implies that a halo must have collapsed by this redshift within the
comoving volume probed by the SDSS survey to find SDSS 1044-1215.
This comoving volume is approximately 5 Gpc$^3$, enclosed by a solid
angle of 600 deg$^2$ and a redshift bin $\Delta z=1$ at redshift
$z=5.8$ in our $\Lambda$CDM cosmology. The appropriate choice for the
size of the redshift bin depends on the parameters of the set of
broad-band filters used in the SDSS. In particular, the Sloan
$z^\prime$ filter that is used to find high-redshift quasars is
centered at 9130\AA with a width of 1900\AA, corresponding to
sensitivity over the redshift interval $5.3 < z < 6.3$. In
Figure~\ref{fig:sdss}, we show the cumulative mass functions, i.e.,
the abundance of all halos with masses exceeding $M$, at $z=20$ and
$z=15$. The dashed curves show the mass function in the $\Lambda$CDM
cosmology, while the solid curves correspond to mass functions in
$\Lambda$WDM models with $m_X=1.25, 1.00, 0.75$, and $0.50$ keV,
respectively from top to bottom. In each panel of
Figure~\ref{fig:sdss}, the horizontal dotted line corresponds to the
abundance of one halo per 5 Gpc$^3$, the requirement for a halo to
host a seed for SDSS 1044-1215.  As the figure shows, for a seed mass
of $M_{\rm seed}=100~{\rm M_\odot}$, the required halo abundance at
$z=15$ results in the limit $m_X\ga 0.47$ keV (or $\vzrms \la 0.119$
km/s). Similarly, for a seed mass of $M_{\rm seed}=10~{\rm M_\odot}$,
the required halo abundance at $z=20$ yields the constraint $m_X\ga
0.79$ keV (or $\vzrms \la 0.059$ km/s).  These limits would strengthen
with the discovery of more than one bright quasar in the $5.3 < z <
6.3$ redshift range. E.g., we find that a total of four objects,
similar in brightness to SDSS 1044-1215, would imply $m_X \ga 0.51$
and $m_X \ga 0.85$ for the high and low seed mass, respectively.

These constraints are somewhat weaker than those we find from
reionization below. Nevertheless, they are comparable to the existing
limits summarized in the Introduction. This type of constraint would
strengthen significantly if the time-averaged mass accretion rate of
BHs was low. In the above estimates, we assumed that the fueling rate
is sufficient to continuously maintain a luminosity near the Eddington
luminosity (and that the growth of the holes is limited by this
luminosity). However, as suggested by \citet{ciot97,ciot00}, this may
not be the case, if the accretion is episodic because the inflow of
gas is quenched by heating of the surrounding gas due to Compton
scattering of the high-energy photons emitted by the central
quasar. On the other hand, a lower radiative efficiency would imply
that the accretion rate could be higher than we have assumed, without
exceeding the Eddington luminosity. However, by comparing the total
mass in BHs at the centers of nearby galaxies with the total light
output of quasars, we know that the average radiative efficiency of
quasars over their lifetimes cannot be much lower than $10\%$, at
least statistically, for the quasar population as a whole
\citep{soltan,barger}. We note further that the simple limit on the
halo abundance we derived here can likely be improved, since not all
halos present at $z=20$ or $z=15$ can plausibly host a seed for the BH
in SDSS 1044-1215.  Indeed, the host-halo of the SDSS quasar is likely
very massive ($M\gg 10^{10}~{\rm M_\odot}$), and among the
high-redshift halos, only those that merge into such large halos by
$z=5.8$ can serve as hosts for the seed BH. A full treatment of this
problem requires the knowledge of the conditional mass function in
$\Lambda$WDM cosmologies, and this is not investigated further in the
present work (but see \citealt{hl00} for a parallel discussion in
usual $\Lambda$CDM cosmologies).

\subsection{Constraints from Reionization.}

\label{ReiCons}

\begin{figure} 
\plotone{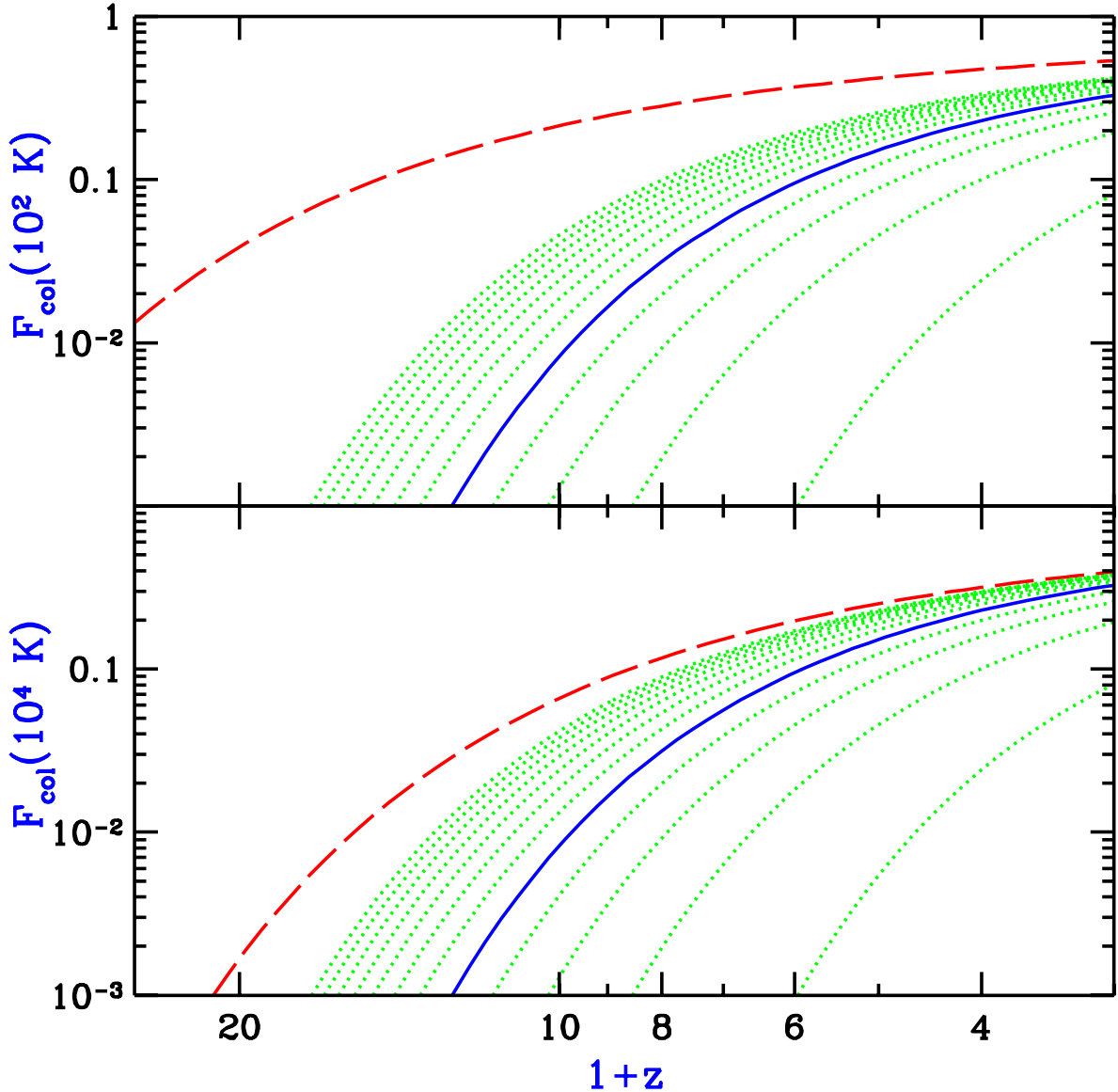} 
\caption{The fraction of baryons collapsed in halos, as a function of 
redshift $z$, for different WDM particle masses. The bottom panel
assumes a minimum halo virial temperature of $10^4$K, which is our
standard case and corresponds to efficient atomic cooling; the top
panel shows the case of a virial temperature of $10^2$K, which
corresponds to efficient cooling in the presence of a significant
abundance of molecular hydrogen. In both panels, the dotted curves
show WDM particles with masses between 0.25 keV and 3 keV, in
increments of 0.25 keV (bottom to top). The $m_X=1.25$~keV case (which
is close to the limit we derive below) is highlighted with a solid
curve. The dashed curve shows the collapsed fraction in a $\Lambda$CDM
universe. Comparing the two panels shows that the reionization
histories in the $\Lambda$CDM models are sensitive to the assumed
minimum cutoff mass, but this sensitivity disappears in the
$\Lambda$WDM models.
\label{fig:fcol}}  
\end{figure}

In this section, we derive the reionization history in universes
dominated by WDM with various particle masses, given the approximate
theory presented in \S~3 above. The main differences in the
reionization histories arise simply from the different collapsed
baryon fractions $F_{\rm col}$ when different particle masses are
assumed. In the bottom panel of Figure~\ref{fig:fcol} we show the
collapsed fraction when the assumed WDM particle mass is between 0.25
keV and 3 keV, in increments of 0.25 keV (dotted curves, bottom to
top). The solid curve highlights the particular case of
$m_X=1.25$~keV, which is near the limit from reionization with our
standard assumptions (see below). The dashed curve shows the collapsed
fraction in a $\Lambda$CDM universe. In each case, $F_{\rm col}$
declines sharply at high redshifts; lower particle masses result in
more significant suppression of the collapsed fraction. While the
bottom panel of Figure~\ref{fig:fcol} assumes our standard value of
$10^4$K for the minimum halo virial temperature required for cooling,
the top panel compares the case of a minimum virial temperature of
$10^2$K, which corresponds to efficient cooling with molecular
hydrogen. Clearly, although the precise cooling threshold is crucial
for $\Lambda$CDM, the halo suppression due to WDM makes halos below a
virial temperature of $10^4$K unimportant for the entire range of
$\Lambda$WDM models which we consider. This is especially true at the
highest redshifts, where (for a fixed virial temperature) the minimum
cooling mass is relatively small.

\begin{figure} 
\plotone{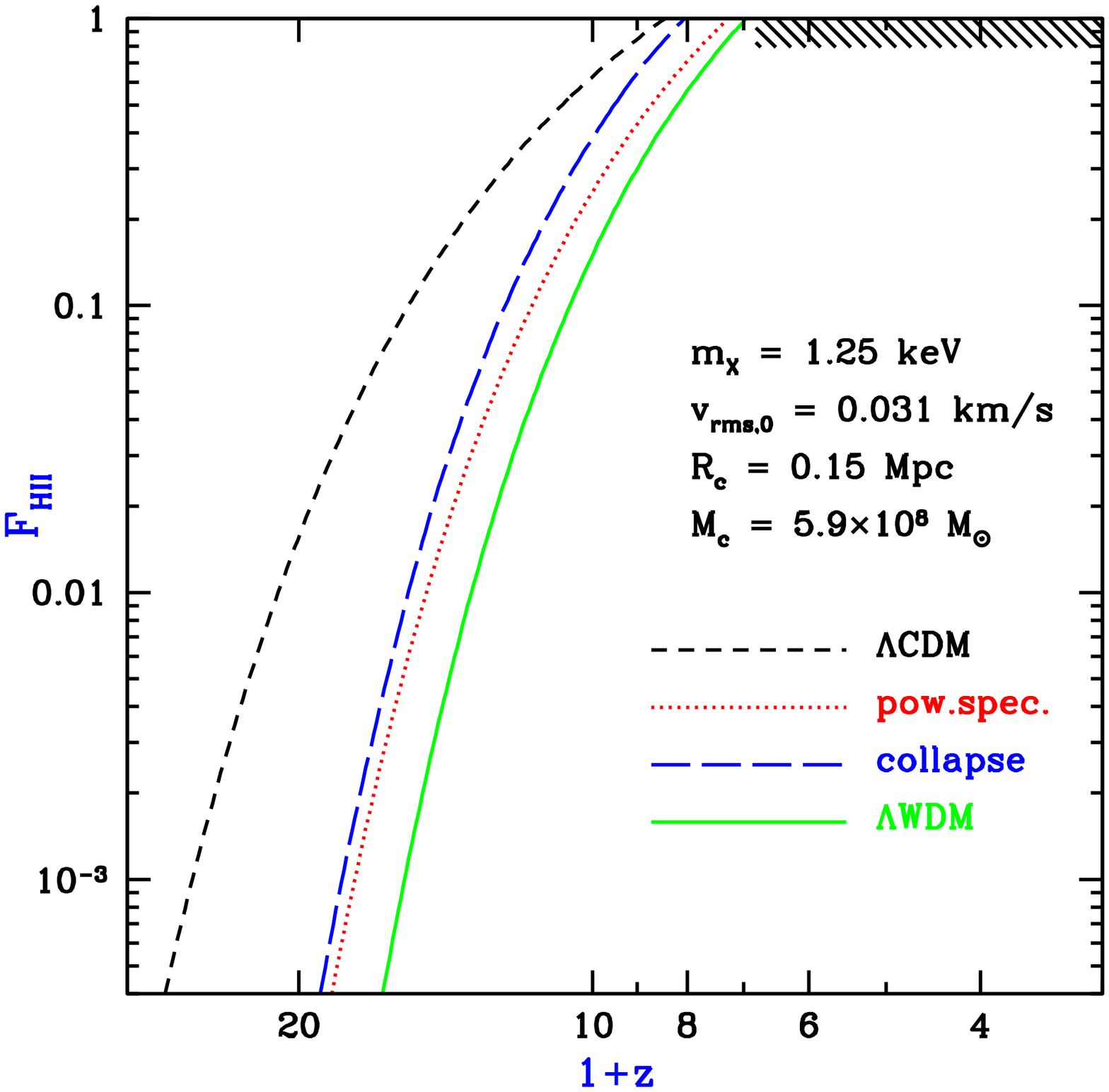} 
\caption{The filling factor of ionized hydrogen, $F_{\rm H\,II}$, as 
a function of redshift $z$ in our standard model ($C=10,
\epsilon_*=0.01$). The uppermost (short dashed) curve corresponds to
$\Lambda$CDM, and the lowest (solid) curve to a WDM particle mass of
$m_X=1.25$ keV. The middle pair of curves shows the separate
contributions to the delay in reionization from the delay in the
collapse of individual halos (long dashed curve) and from the
suppression of the power spectrum (dotted curve). The shaded region in
the upper right corner indicates the requirement $z_{\rm
reion}>5.8$. Also given in the figure are the values of the
corresponding velocity $\vzrms$ as well as the cutoff scale $R_c$ and
mass $M_c$, defined using the $z=0$ power spectrum.
\label{fig:reion}}
\end{figure}

Once the collapsed fraction is known in a given cosmology, the filling
factor of ionized regions, $F_{\rm H\,II}$ follows as discussed in \S
\ref{sec:reion} above.  As an example, in Figure~\ref{fig:reion}, we
show the evolution of $F_{\rm H\,II}$ in our standard model ($C=10,
\epsilon_*=0.01$), assuming a WDM particle mass of 1.25 keV ($\vzrms =
0.031$ km/s; solid curve). For reference, the short-dashed curve shows
the filling factor in the $\Lambda$CDM model.  As discussed above, WDM
reduces the collapsed fraction of baryons due to two distinct reasons:
first because the power spectrum is suppressed at low masses; and
second, because the collapse of each low-mass halo near and below the
critical cutoff mass is delayed.  In order to assess these effects
separately, in Figure~\ref{fig:reion} we show the evolution of the
filling factor when the power spectrum is suppressed, but the collapse
of each halo is kept the same as in $\Lambda$CDM (dotted curve); and
conversely, when the collapse of each halo is delayed by the WDM, but
the power spectrum is kept the same as in $\Lambda$CDM (long-dashed
curve).  This pair of curves reveals that both effects lead to a
substantial reduction of the total amount of collapsed gas. The two
effects are comparable at high redshift, but the power spectrum cutoff
is dominant at reionization. When both effects are included, the
filling factor reaches a value of unity only at a redshift of $z=6.0$,
as opposed to a redshift of $z\approx 7.4$ in the corresponding
$\Lambda$CDM case.

The generic trend demonstrated in Figure~\ref{fig:reion} holds when
different particle masses are assumed: the delayed collapse of
individual halos, and the suppression of the power spectrum both
contribute to a delay in the reionization epoch, with the latter
effect dominating. As can be expected from Figure~\ref{fig:fcol}, the
reionization redshift is a strong function of the assumed WDM particle
mass. The current best lower limit on the reionization redshift is
$z=5.8$, inferred from the spectrum of a bright quasar found at this
redshift in the Sloan Digital Sky Survey \citep[SDSS,][]{f00}. In
our models, we define reionization to occur at the redshift when the
filling factor formally reaches unity.  Although this choice is
somewhat arbitrary, it is likely a conservative assumption, since even
after the entire IGM is reionized, its opacity remains very large for
some time until the neutral hydrogen fraction drops to $\la
10^{-6}$. We compute the evolution of $F_{\rm H\,II}$ for various WDM
masses $m_X$ and obtain the reionization redshift under this
definition.

\begin{figure} 
\plotone{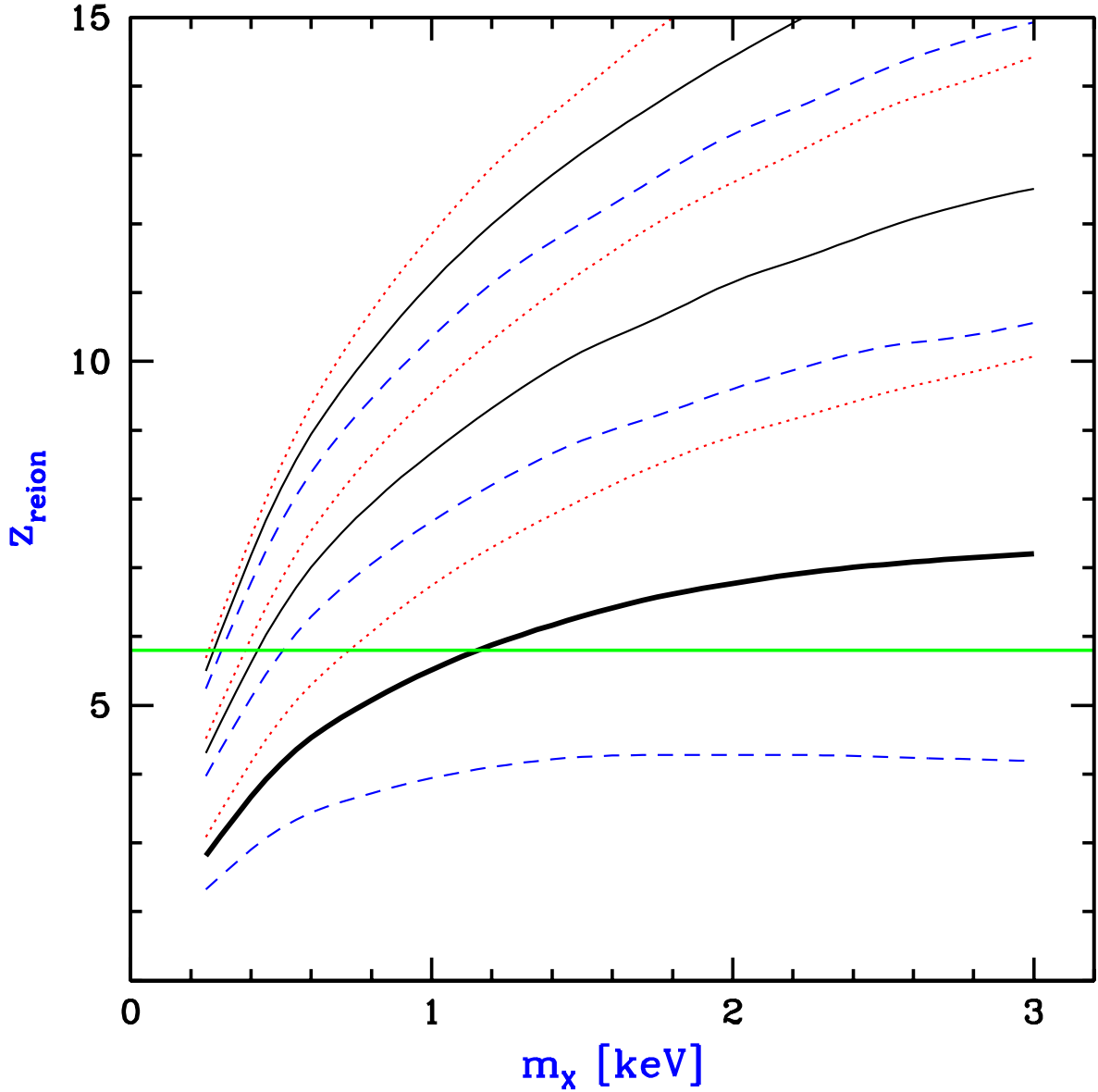} 
\caption{Reionization redshift as a function of WDM particle mass. 
Our standard model, which adopts standard values for the
star-formation efficiency, escape fraction, and stellar IMF, is shown
by the thick solid line.  The three solid curves show models with the
same clumping factor ($C=10$) but different ionization efficiencies
($\epsilon_*=0.01,0.1,1$, bottom to top).  The three dotted curves
correspond to the same models except with $C=1$; similarly, the three
dashed curves correspond to these models except with $C=30$.  In our
standard model, the requirement that $z_{\rm reion}>5.8$ limits the
WDM particle mass to be $E > 1.2$~keV; in the most extreme model, the
limit is lowered to $E\ga 0.25$~keV.
\label{fig:zreion}}
\end{figure}

Figure~\ref{fig:zreion} shows the resulting value of $z_{\rm reion}$
as a function of WDM particle mass in our models. The thick solid
curve corresponds to our standard model, and results in a limit of 1.2
keV on the WDM particle mass (which corresponds to $\vzrms \leq 0.033$
km/s). In order to assess the robustness of the limit on the WDM
particle mass, in Figure~\ref{fig:zreion} we also show results for
variations away from our standard parameters values. The top and
middle solid curves assume $\epsilon_*=1$ and $\epsilon_*=0.1$,
respectively, and yield limits of 0.28 and 0.43 keV.  For each of the
models represented by the three solid curves (which all assume a
clumping factor $C=10$), we also show corresponding results when $C=1$
(upper, dotted curve), and when $C=30$ (lower, dashed curve). We
regard the upper-most triplet of models as extreme cases, where the
star formation efficiency and escape fraction are both adjusted to
their highest possible values. An exception to this is the stellar
IMF: if it is heavily biased towards high-mass stars, the ionizing
photon yields can be substantially increased.  Note, however, that
even in the most extreme models, the constraint $E>0.25$ keV follows.

An interesting feature of Figure~\ref{fig:zreion} is the strong
dependence of the limiting particle mass on the assumed limit on the
reionization redshift. Although the redshift limit is currently
$z=5.8$, if $\Lambda$CDM is correct, future observations could soon
increase this redshift and tighten the constraint on WDM. If the limit
is increased beyond $z=7.4$, it would rule out the model with our
standard parameters, even in $\Lambda$CDM. The {\it Next Generation
Space Telescope}\, (\NGST\,) could push the redshift limit to $z>10$,
which would greatly constrain $\Lambda$WDM even with reionization
parameters stretched to values that appear unlikely at present. Also,
models of cosmological parameter estimation from future CMB
measurements \citep{eht99, zss97} suggest that the reionization
redshift should be easily measurable with the Planck satellite, and
the polarization signature of reionization is likely detectable even
with the upcoming MAP satellite.

It is useful to assess the individual contributions of the power
spectrum cutoff, and delayed collapse of halos, to these limits on the
WDM particle mass.  As an example, in models with a constant clumping
of $C=10$, Figure~\ref{fig:zreion} reveals the limits of 0.28, 0.43,
and 1.17 keV for $\epsilon_*=1$, 0.1, and 0.01, respectively. If the
power spectrum cutoff were ignored, these limits would change to 0.17,
0.29, and 0.71 keV; similarly, if the effect of the WDM on the
collapse of halos were ignored, the limits would be 0.15, 0.29, and
0.98 keV. As before, the cutoff plays a larger role than halo
dynamics. Note that the constraints that we obtain when only the
cutoff is included would also apply to any modification of the
$\Lambda$CDM model in which the power spectrum is sharply cut off
below the scale $R_c$ given by equation (\ref{eq:Rc}).

As mentioned in \S \ref{sec:reion}, we can normalize the parameter
$\epsilon_*$ in our models by computing the background flux produced
at redshift $z=3$ and comparing with observations. The measured value
at $z \sim 3$ from the proximity effect is around $J_{21}=0.7$, i.e.,
the proper intensity is $J=0.7\times 10^{-21}~{\rm erg~s^{-1}~
cm^{-2}~Hz^{-1}~sr^{-1}}$ (\citealt{prox1}; see also \citealt{prox2}
for the basic measurement method). The observed flux at redshift
$z<z_{\rm reion}$ and frequency $\nu$, in units of ${\rm
erg~cm^{-2}~s^{-1}~Hz^{-1}~sr^{-1}}$ is given by
\citep[e.g.,][]{Peebles93}
\begin{equation}
J(\nu,z)=\frac{c}{4\pi}\int_z^{z_{\rm reion}} dz^\prime \frac{dt}
{dz^\prime} \, j^*_{\nu_z^\prime}(z^\prime) \left( \frac{1+z}
{1+z^\prime}\right)^3 \exp \{-\tau(z,z^\prime,\nu)\},
\label{eq:6}
\end{equation}
where $c \, dt/dz$ is the cosmological line element per unit redshift,
$\nu_z^\prime = \nu(1+z^\prime)/(1+z)$ is the frequency appropriate
for photons emitted at redshift $z^\prime$, $j^*$ is the proper
emission coefficient in units of ${\rm erg~cm^{-3}~s^{-1}~Hz^{-1}}$,
and $\tau(z,z^\prime,\nu)$ is the effective optical depth between
redshifts $z$ and $z^\prime$ for a photon emitted at frequency
$\nu_z^\prime$ at $z^\prime$.  Note that we impose the upper limit
$z_{\rm reion}$ on the integral, since photons emitted prior to this
redshift are used to reionize the universe, and do not contribute to
the background flux. \citet{Lyopacity} calculated the optical depth
$\tau(z,z^\prime,\nu)$ based on the distribution of Ly$\alpha$
absorption systems in redshift and hydrogen column density; we adopt
this optical depth but with a factor of 0.6 correction, for
consistency both with more recent observations of the distribution of
Ly$\alpha$ absorption systems \citep{fardal} and with a recent
measurement of absorption in quasar spectra at $z=3.4$
\citep{steidel}. In our models, the stellar emission coefficient is
given by a sum of the instantaneous emission from all the stars (of
different ages) present at $z$,
\begin{equation}
j^*_{\nu}(z)=\epsilon_*\Omega_{\rm b}\rho_{\rm
c0}(1+z)^3
\int_z^{\infty} dz^{\prime}\frac{dF_{\rm col}}{dz^{\prime}}
\epsilon^{\rm star\,}[\nu(1+z^{\prime}),t_{z,z^{\prime}}],
\label{eq:7}
\eeq where $\rho_{\rm c0}$ is the current critical density of the
universe, $t_{z,z^{\prime}}$ is the time interval between the redshift
$z^{\prime}$ at which a star was born, and the redshift $z$, and
$\epsilon^{\rm star\,}(\nu,t)$ is the composite physical emissivity
(in ${\rm erg~s^{-1}~Hz^{-1}~M_{\odot}^{-1}}$) of a population of
stars with our adopted IMF \citep{IMF}, as given by the \citet{bc96}
model at time $t$ after an initial starburst. In calculating $dF_{\rm
col}/dz^{\prime}$ in equation (\ref{eq:7}) we include the suppression
of the formation of dwarf galaxies after reionization
\citep[e.g.,][]{efstathiou, thoul, navarro}, approximated here as a
sharp cutoff at a halo circular velocity of 40~km/s. The severe
opacity faced by ionizing photons from high redshifts implies that
$J_{21}$ at $z=3$ essentially fixes the star formation rate at
$z=3$. For different WDM masses $m_X$, all with the same efficiency
$\epsilon_*$, the cosmic star formation rate differs greatly at high
redshift but by $z=3$ is fairly insensitive to $m_X$.

The measured value of $J_{21}$ is highly uncertain and may be about a
factor of 2 higher or lower. It is also, in fact, averaged over
$z=2$--4. In applying this measurement to our models, we must consider
the possibility that different source populations dominate the
ionizing intensity at $z \sim 3$ and at reionization. Indeed, a
decrease with redshift in the contribution of quasars to the ionizing
background around $z \sim 3$ is suggested by observations of variable
\ion{He}{2} opacity \citep[e.g.,][]{heap, anderson}, evolution in the
\ion{Si}{4}/\ion{C}{4} ratio in Ly$\alpha$ absorption systems
\citep{songaila, boksenberg}, and the relatively high temperature of
the IGM at $z\sim3$ \citep{ricotti, schaye, bryan, mcdonald}. If the
trend indicated by these observations continues to higher redshift, it
justifies our reionization models which include only stellar
emission. However, the observed $J_{21}$ at $z=2$--4 may include a
significant contribution from quasars. Indeed, the value of $J_{21}$
derived from the proximity effect is consistent with the ionizing flux
expected from quasars \citep{haardt}. Thus, our $m_X=1.2$ keV model
with standard parameters, which reionizes at $z=5.8$, is fully
consistent with observations of the proximity effect; in the simple
calculation based on equations (\ref{eq:6}) \& (\ref{eq:7}) above it
yields a flux of $J_{21}=0.36$ at $z=3$, consistent with the measured
value of $J_{21}=0.7$ if quasars and stars make a comparable
contribution to the observed $J_{21}$. Given the uncertainties,
however, we estimate a range of possible normalizations of our model
by allowing a stellar contribution to $J_{21}$ that is higher or lower
by a factor of 2. At the lower end, if $J_{21} < 0.22$ then even the
$\Lambda$CDM model does not reionize by $z=5.8$. At the higher end, if
stars contribute a $J_{21}=0.72$ at $z=3$ then our limit on WDM
weakens to $m_X > 0.75$ keV.

Although normalizing our models to a fixed value of $J_{21}$ adds a
direct observational constraint, the computation of the flux in our
models depends on the assumption that $\epsilon_*$ does not evolve
with redshift. We also note here an observational result which may
conflict with those mentioned above. \citet{steidel} have reported a
preliminary detection of significant Lyman continuum flux in a
composite spectrum of 29 Lyman break galaxies at $z=3.4$. The observed
flux level implies a large escape fraction of ionizing radiation, and
may be inconsistent with the spectral break at 912\AA\ expected from
stars with the IMF that our models assume. If the \citet{steidel}
results are representative of the galaxy population as a whole, then
galaxies contribute $\sim 5$ times more ionizing flux than quasars to
the background flux at $z=3.4$. On the other hand, \citet{heckman}
applied a different method based on interstellar absorption lines and
found a very low $f_{\rm esc}<6\%$ in several local starburst galaxies
and one such galaxy at $z=2.7$. Further measurements are needed to
settle this issue, especially cross-checks where different methods are
applied to the same galaxies. In addition, measurements of the
probability distribution of transmitted flux in the Lyman-$\alpha$
forest of quasar absorption lines also suggest a low ionizing
background, $J_{21} \sim 0.3$ \citep{rauch,mcdonaldA}, although this
method for constraining $J_{21}$ is less direct than the proximity
effect as it relies on an understanding of the thermal state of the
Lyman-$\alpha$ forest.

Another observational constraint on our models is the metallicity of
the IGM at $z=3$. For the \citet{IMF} IMF, if we assume that a
supernova is produced by each $M>8 M_{\odot}$ star, then on average
one supernova explodes for every 126 $M_{\odot}$ of star formation,
expelling $\sim 1\, M_{\odot}$ of heavy elements. Hydrodynamic
simulations by \citet{maclow} suggest that the hot, metal-enriched
ejecta tends to escape from the small halos which typically host
galaxies at high redshift. If on average half the metals are expelled
and are effectively mixed into the IGM, the resulting average
metallicity of the IGM is \beq Z_{\rm IGM} = 0.02\, \left(\frac{f_*}
{0.1}\right)\, F_{\rm col}\, Z_{\odot}\ , \eeq where $f_*$ is the star
formation efficiency as in \S \ref{sec:reion}, and $Z_\odot=0.019$ is
the solar metallicity. For the $m_X=1.2$ keV model at $z=3$, this
yields $Z_{\rm IGM} \sim 4 \times 10^{-3} Z_{\odot}$. This is
consistent with the value of $Z_{\rm IGM}\sim 10^{-2.5 (\pm
0.5)}Z_\odot$ \citep[e.g.,][]{t95, l98, c98, e00, p00} observed in low
column density Lyman-$\alpha$ absorbers, assuming that the metallicity
of these systems is representative of the average metallicity of the
IGM.

Despite the degeneracy in the dependence of $z_{\rm reion}$ on $m_X$
and on $\epsilon_*$, the value of each parameter can be isolated via
an independent consistency check which will be easily accessible to
\NGST.\, The key observation is the luminosity function of galaxies at
redshifts around reionization. As shown above, the effect of WDM is to
produce a sharp low-mass cutoff in the halo mass function. For an
$m_X$ value within the range considered above, the resulting galaxy
luminosity function is cut off at luminosities at which \NGST\, would
otherwise detect large numbers of galaxies. To estimate the galaxy
luminosity function, we assume a star formation efficiency of
$f_*=0.1$ and also that the stars in each galaxy at a given $z$ had
formed continuously during a time equal to $20\%$ of the age of the
universe at $z$. We assume a metallicity $Z=0.001$, and use the
stellar population model results of \citet{starb}, with a Ly$\alpha$
cutoff in the spectrum due to absorption by neutral hydrogen.

We show the resulting luminosity functions of galaxies in
Figure~\ref{fig:LumFn}. The predicted luminosity function (in the
\NGST\, wavelength band) is shown at $z=8$ (solid curves) and $z=12$
(dashed curves), in each case assuming that reionization has not yet
occurred by the plotted redshift. At each redshift, the four curves
correspond, from top to bottom, to $\Lambda$CDM, and $\Lambda$WDM with
$m_X=1$ keV, 0.75 keV, and 0.5 keV. Also shown is the detection
threshold for \NGST\, (vertical dotted line). The figure clearly shows
that any WDM model with $m_X \la 1$ keV ($\vzrms \ga 0.041$ km/s)
produces a cutoff which is detectable with \NGST\, for any $z_{\rm
reion} \sim 10$. Note that a change in the value of $\epsilon_*$
simply shifts each curve right or left but does not affect the overall
shape. In general, it may be easier to detect a cutoff due to WDM at
high redshift, because the formation of additional dark matter halos
due to fragmentation should be less significant than at low redshift,
although in any case these additional halos should have masses
significantly lower than the mass of the characteristic bend 
(see \S \ref{sec:Abundance}).  The cutoff in the luminosity function
can also be detected at $z < z_{\rm reion}$, since although the
increased gas pressure in ionized regions suppresses the formation of
dwarf galaxies, the resulting cutoff is expected to occur at lower
fluxes than for WDM and the cutoff is not nearly as sharp \citep{me}.

\begin{figure} 
\plotone{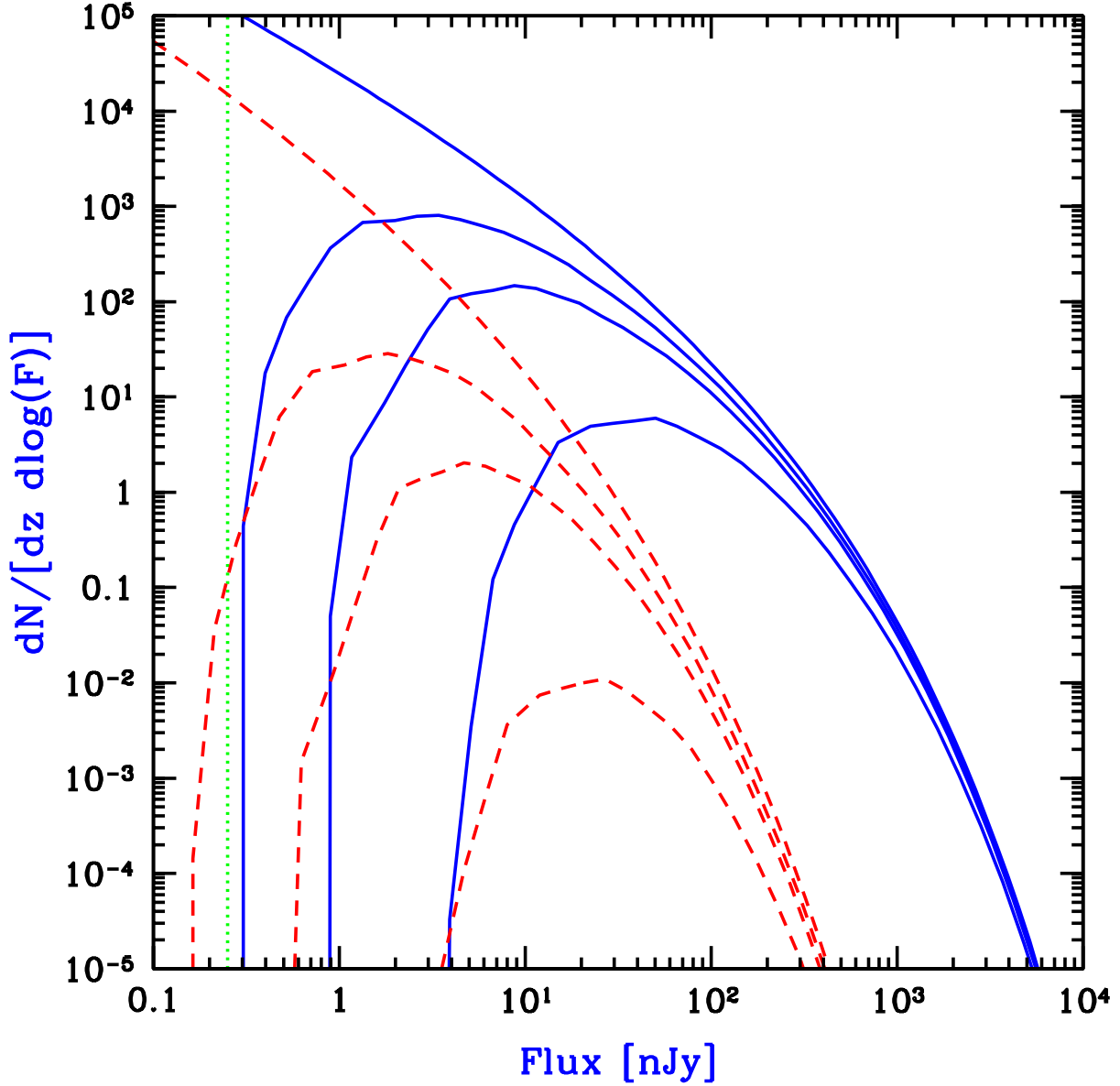} 
\caption{Galaxy luminosity functions in WDM. The distribution 
$dN/[dz\,d\log(F)]$ is shown at $z=8$ (solid curves) and $z=12$
(dashed curves), where $F$ is the mean flux in the \NGST\, wavelength
band (0.6--3.5$\mu$m), and $dN/dF$ is the number of galaxies per
\NGST\, field of view ($4\arcmin \times 4\arcmin$) and per unit 
flux. At each redshift, the top curve corresponds to $\Lambda$CDM, and
the other curves to $\Lambda$WDM with $m_X=1$ keV, $m_X=0.75$ keV, and
$m_X=0.5$ keV, respectively from top to bottom. Also shown is the
detection threshold for \NGST,\, i.e., the minimum flux at which a
source can be detected at 5-$\sigma$ in a long exposure (vertical
dotted line).
\label{fig:LumFn}}
\end{figure}
	
Finally, we consider, within $\Lambda$CDM, changes in the primordial
power spectrum index $n$ away from its fiducial value of
$n=1$. Inflationary models allow $n$ to be as small as $n\approx 0.7$
\citep{ll92}; lowering the value of $n$ reduces small scale power, and
thus delays the reionization epoch. In our standard $\Lambda$CDM model
with $\sigma_8$ fixed at 0.9, we find that the reionization redshift
of $z=5.8$ translates to a limit $n>0.80\, (C=10,\epsilon_*=0.01)$ or
$n>0.46 \, (C=10, \epsilon_*=0.10)$.

\section{Conclusions}

In this paper, we have studied constraints on WDM-dominated
cosmological models based on the high-redshift universe. Although we
have focused on WDM models, our constraints should apply generically
to any model in which the primordial power-spectrum is suppressed on
small scales relative to its amplitude in CDM cosmologies. The loss of
small-scale power reduces the number of collapsed objects at high
redshifts ($z>5$), and makes it more difficult to reionize the
universe, or to grow a single supermassive BH with the mass inferred
to reside in the quasar SDSS 1044-1215 discovered in the Sloan Digital
Sky Survey at $z=5.8$.

We have quantified these constraints in WDM models, utilizing a halo
mass function derived by an extension of the excursion set formalism.
We have found that in addition to the loss of small-scale power, there
is a delay in the nonlinear collapse dynamics of WDM halos, caused by
the particle velocities. Unlike the suppression of small-scale power,
this phenomenon is specific to WDM models, and is comparable in
importance to the loss of small-scale power (though its overall effect
on the reionization redshift is somewhat smaller). Also, we showed
that due to the adiabatic decay of the WDM velocity dispersion, most
of the delay in the collapse dynamics occurs at very high
redshifts. This effect is naturally included in N-body simulations,
which typically begin at a lower redshift but use the correct initial
power spectrum and a velocity distribution which is generated
self-consistently from the density distribution. Our equation
(\ref{MJeans}) for the Jeans mass implies that the additional effects
of random velocities at $z \la 40$ are negligible. We noted a certain
ambiguity in the semi-analytic models, in that the overall suppression
depends somewhat on the starting redshift, but we obtained a good
match to the results of N-body simulations by starting at
matter-radiation equality.

We have found that if high-redshift galaxies produce ionizing photons
with an efficiency similar to their $z=3$ counterparts, reionization
by redshift $z=5.8$ places a limit of $m_X\ga 1.2~$ keV ($\vzrms \la
0.033$ km/s) on the mass of the WDM particles. This limit is somewhat
stronger than the limit inferred from the statistics of the Ly$\alpha$
forest (which yields $m_X\ga 0.75~$keV; \citealt{vijay00}), although
our limit may weaken to $m_X\ga0.75~$keV ($\vzrms = 0.060$ km/s) given
the uncertainty in current measurements of the stellar contribution to
the ionizing intensity at $z=3$. If we relax the assumption of a
constant $\epsilon_*$ (and stellar IMF) with redshift, our limit
weakens to $m_X\ga0.4~$keV ($\vzrms \la 0.14$ km/s) if the
ionizing-photon production efficiency is ten times greater at $z > 6$
than it is at $z \sim 3$. In our analysis, we have assumed a universal
efficiency of ionizing-photon production in WDM halos. If, on the
other hand, this efficiency declines in low-mass halos (due to
feedback from supernovae), this will tighten our constraints. We have
also shown that WDM models with $m_X\la1~$keV ($\vzrms \ga 0.041$
km/s) produce a low-luminosity cutoff in the high-$z$ galaxy
luminosity function which is detectable with the {\it Next Generation
Space Telescope}\,; such an observation would directly break the
degeneracy in the reionization redshift between a low ionizing-photon
production efficiency and a small WDM particle mass. Our results also
imply that the existence of a $\approx 4\times 10^9~{\rm M_\odot}$
supermassive black hole at $z=5.8$, believed to power the quasar SDSS
1044-1215, yields the somewhat weaker, but independent limit $m_X\ga
0.5$ keV (or $\vzrms \la 0.10$ km/s).

In summary, at present, our work leaves open the possibility that WDM
consists of particles with a mass near $\sim 1$ keV. The various
constraints derived here should tighten considerably as observations
probe still higher redshifts. If future observations uncover massive
black holes at $z \ga 10$ or reveal that reionization occurred at
$z\ga 10$, this would conclusively rule out WDM models as the solution
to the current small-scale crisis of the CDM paradigm.

\acknowledgments

We are grateful to Chung-Pei Ma and to Neil Turok for values of the
WDM power spectrum at equality. We thank Paul Bode for the values of
halo abundances in WDM simulations. We thank the ITP (Santa Barbara)
where the ideas for this paper were conceived during the Spring 2000
program on ``Galaxy Formation and Evolution''. ZH also thanks
Joel Primack for independently suggesting that reionization places
constraints on WDM models. RB acknowledges support from CITA and from
Institute Funds (IAS, Princeton). This research is supported in part
by NASA through the Hubble Fellowship grant HF-01119.01-99A, awarded
to ZH by the Space Telescope Science Institute, which is operated by
the Association of Universities for Research in Astronomy, Inc., for
NASA under contract NAS 5-26555.

\end{document}